%% file: alp_lfv_belleII.tex
\documentclass[11pt,a4paper]{article}

\pdfoutput=1
\usepackage{jheppub}
\usepackage{graphicx,color,float}
\usepackage{hyperref}
\usepackage{multirow}
\usepackage{verbatim}
\usepackage{longtable}
\usepackage{amsmath}
\usepackage{amsfonts}
\usepackage{amssymb}
\usepackage{rotating}

\usepackage[normalem]{ulem}

\def\gsim{\raise0.3ex\hbox{$\;>$\kern-0.75em\raise-1.1ex\hbox{$\sim\;$}}}
\def\lsim{\raise0.3ex\hbox{$\;<$\kern-0.75em\raise-1.1ex\hbox{$\sim\;$}}}

\begin{document}


\title{Probing charged lepton flavor violation with axion-like particles at Belle II}

\author[a,b,c]{Kingman Cheung}
\emailAdd{cheung@phys.nthu.edu.tw}
\affiliation[a]{Department of Physics, National Tsing Hua University,	Hsinchu 300, Taiwan}
\affiliation[b]{Center for Theory and Computation, National Tsing Hua University,	Hsinchu 300, Taiwan}
\affiliation[c]{Division of Quantum Phases and Devices, School of Physics, Konkuk University, Seoul 143-701, Republic of Korea}

\author[d]{Abner Soffer}
\emailAdd{asoffer@tau.ac.il}
\affiliation[d]{School of Physics and Astronomy, Tel Aviv University, Tel Aviv 69978, Israel}

\author[a,b]{Zeren Simon Wang}
\emailAdd{wzs@mx.nthu.edu.tw}

\author[a]{Yu-Heng Wu}
\emailAdd{yuheng1k996@gapp.nthu.edu.tw}

\date{\today}


\vskip1mm

\abstract{
We study charged lepton flavor violation associated with a light leptophilic axion-like particle (ALP), $X$, at the $B$-factory experiment Belle II.
We focus on production of the ALP in the tau decays $\tau \to  X l$ with $l=e,\mu$, followed by its decay via $X\to l^- l^+$.
The ALP can be either promptly decaying or long-lived.
We perform Monte-Carlo simulations, recasting a prompt search at Belle for lepton-flavor-violating $\tau$ decays, and propose a displaced-vertex (DV) search. For both types of searches, we derive the Belle II sensitivity reaches in both the product of branching fractions and the ALP coupling constants, as functions of the ALP mass and lifetime.
The results show that the DV search exceeds the sensitivity reach of the prompt search to the relevant branching fractions by up to about a factor of 40 in the long decay length regime.
}
  



\maketitle

%


\input{tex/intro}

\input{tex/model}

\input{tex/constraints}

\input{tex/simulation}

\input{tex/results}
\input{tex/conclusions}

\section*{Acknowledgment}

We thank Martin Hirsch for useful discussions.
Z.S.W. is supported by the Ministry of Science and Technology (MoST) of Taiwan with grant numbers MoST-109-2811-M-007-509 and MoST-110-2811-M-007-542-MY3.
K.C. and Y.W. were supported by MoST-107-2112-M-007-029-MY3 and
MoST-110-2112-M-007-017-MY3.
AS is supported by grants from the Israel Science Foundation, the US-Israel Binational Science Fund, the Israel Ministry of Science, and the Tel Aviv University Center for AI and Data Science.
\bibliographystyle{JHEP}
\bibliography{bib}

\end{document}

%% file: tex/intro.tex

\section{Introduction}\label{sec:intro}

The latest experimental bounds on the neutron electric dipole moment have constrained the strength of
Quantum Chromodynamics (QCD)  $\theta$-term in the Standard Model (SM) of particle physics to be
less than $\sim 10^{-10}$~\cite{nEDM:2020crw,Dragos:2019oxn}, pointing to a fine-tuning issue,
known as the strong CP problem in QCD~\cite{Peccei:1977ur,Peccei:2006as,ParticleDataGroup:2020ssz}.     
One of the most compelling solutions to this problem is to introduce a new global symmetry $U(1)_{\text{PQ}}$, which is spontaneously broken by a dynamical CP-conserving axion field~\cite{Peccei:1977ur,Peccei:2006as}.
Doing so predicts a new pseudoscalar boson, the pseudo-Nambu-Goldstone boson of the broken $U(1)_{\text{PQ}}$ symmetry, called the QCD axion, $a$.
The breaking scale of the new symmetry has to be much higher than the electroweak scale: $f_a \gtrsim 10^{9}\text{ GeV}$~\cite{Feng:1997tn}.
Since the mass of the QCD axion, $m_a$, as well as its couplings to the SM particles, are both inversely proportional to $f_a$, the QCD axion should be very light ($10^{-6}\text{ eV}< m_a <10^{-2} \text{ eV}$) and feebly interacting, with a very long lifetime~\cite{ParticleDataGroup:2020ssz}.

Similar to the QCD axion, an axion-like particle (ALP), denoted $X$ throughout this work, is a pseudoscalar boson that interacts feebly with SM particles. Unlike the QCD axion, an ALP does not have a linear proportionality relation between its mass $m_X$ and its couplings to the SM particles, and hence does not necessarily solve the strong CP problem.
ALPs can be as light as $10^{-22} \text{eV}$~\cite{Kim:2015yna,DeMartino:2017qsa}
or as heavy as 1 TeV (heavy GUT axions)~\cite{Rubakov:1997vp}.
Axions and ALPs are considered some of the most plausible candidates for dark matter~\cite{Dine:1982ah,Abbott:1982af,Preskill:1982cy,Marsh:2015xka,Lambiase:2018lhs,Auriol:2018ovo,Houston:2018vrf}.

ALPs can be potentially produced at various experiments, including the Large Hadron Collider and $B$-factories
(see for instance Refs.~\cite{Jaeckel:2015jla,Brivio:2017ije,Dolan:2017osp,Bellazzini:2017neg,Bauer:2017ris,Knapen:2017ebd,Bauer:2018uxu,Aloni:2018vki,Carmona:2021seb} for related studies).
Depending on their masses and strengths of the couplings to the SM particles, ALPs can be short-lived and decay promptly, or be long-lived and decay after traveling a macroscopic distance, leading to signatures of displaced objects or missing energy.
ALPs can couple to gauge bosons as well as quarks and leptons, giving rise to a large class of possible experimental signatures.
For instance, using its initial dataset of about 445~pb$^{-1}$, Belle~II has reported the results of a search for a light pseudoscalar boson that couples to photons~\cite{Belle-II:2020jti}.

In this paper, we restrict ourselves to the case in which the ALPs are leptophilic, i.e., couple at tree level only to the charged leptons, and neglect couplings to neutrinos.
In general, the ALP couplings to the charged leptons are not necessarily diagonal, and may have charged-lepton-flavor-violating (LFV) interactions~\cite{Heeck:2016xwg}.
Since the SM predicts no flavor-changing neutral current (FCNC) at tree level, experimental observations of LFV can be a clear signature of new physics.
For muon LFV processes such as $\mu\to e\gamma$ and $\mu\to 3e$, the SM predictions for the branching fractions are about $10^{-55}$~\cite{Petcov:1976ff,Hernandez-Tome:2018fbq}, much lower than the present experimental limits BR$(\mu\to e\gamma)<4.2\times 10^{-13}$~\cite{MEG:2016leq} and BR$(\mu\to 3e)< 1.0\times 10^{-12}$~\cite{SINDRUM:1987nra}.
The future experiments Mu3e and MEG II will have an improved sensitivity to these processes by as much as four orders of magnitude~\cite{Blondel:2013ia,Baldini:2013ke}.
Compared to the muon-related LFV, LFV effects in the $\tau$ sector are much more loosely constrained, as we briefly discuss below.
Studies of FCNCs with the ALPs can be found for the lepton sector in Refs.~\cite{Cordero-Cid:2005vca,Dev:2017ftk,Bauer:2019gfk,Cornella:2019uxs,Calibbi:2020jvd,Endo:2020mev} and for the quark sector in Refs.~\cite{Batell:2009jf,Kamenik:2011vy,Gavela:2019wzg,Carmona:2021seb}.

In order to study the ALP couplings to the $\tau$ lepton,  we consider the ongoing $B$-factory Belle~II experiment~\cite{Belle-II:2010dht,Belle-II:2018jsg} at the SuperKEKB collider, where energy-asymmetric collisions of electron and positron beams at the center-of-mass energy $\sqrt{s}=10.58$~GeV take place.
Belle~II is planned to collect about $4.6\times 10^{10}$ 
$e^+ e^-\to\tau^+\tau^-$ events with the design integrated luminosity of 50~ab$^{-1}$~\cite{HernandezVillanueva:2018dqu}.
The large dataset will enable probing rare $\tau$ decays, including LFV decays, with unprecedented precision.

We propose to search for the decays $\tau\to X l$ with $X\to l^-l^+$, where $l$ can be either an electron or a muon.
This restricts the ALP mass to the range $2\, m_l < m_X < m_\tau - m_l$.
In this mass range, the strongest published limits on lepton-flavor-conserving ALP couplings stem from the beam-dump experiments conducted at SLAC~\cite{Essig:2010gu,Andreas:2010ms} and a dark-photon search performed at BaBar~\cite{BaBar:2016sci,Bauer:2017ris}, both of which assumed no tree-level LFV couplings.
Bounds on the LFV ALP couplings can also be derived from  $\tau\to Xe$, where $X$ is invisible (see studies at ARGUS~\cite{ARGUS:1995bjh}, Belle~\cite{Yoshinobu:2017jti}, and Belle~II~\cite{Guadagnoli:2021fcj,DeLaCruz-Burelo:2020ozf,Ma:2021jkp,Calibbi:2020jvd,Tenchini:2020njf}).
This signature, which is not covered in this work, suffers from large background from $\tau\to l \nu \nu$, requiring more sophisticated methods to suppress. In Section~\ref{subsec:lepton_universality} we extract stronger constraints from this signature based on test of lepton-flavor universality in leptonic $\tau$ decays~\cite{BaBar:2009lyd}, relevant for the case of interest of non-universal couplings.
For past studies on probing LFV associated with an ALP at Belle II, see e.g. Refs.~\cite{Endo:2020mev,Iguro:2020rby}.

The previous Belle experiment~\cite{Belle:2000cnh} has searched for lepton-flavor-violating $\tau$ decays into three charged leptons and obtained limits on the decay branching ratios in the order of magnitude of $10^{-8}$~\cite{Hayasaka:2010np} (see also Ref.~\cite{BaBar:2010axs} for a similar search at the BaBar experiment).
In this work we recast that search into the studied theoretical scenarios of a leptophilic ALP, obtaining tighter constraints than previously published ones.  Assuming equal selection efficiencies at Belle and Belle~II, we calculate the expected Belle~II limits on the model-independent as well as model-dependent parameters.

In the main part of this work, we propose a displaced-vertex (DV) search strategy that targets long-lived ALPs.
Such long-lived ALPs belong to a large class of long-lived particles (LLPs) widely predicted in various beyond-the-SM models.
See Refs.~\cite{Curtin:2018mvb,Lee:2018pag,Alimena:2019zri} for reviews of LLP collider searches and models.
For past studies of LLPs at Belle~II, see, e.g., Refs.~\cite{Dib:2019tuj,Kim:2019xqj,Kang:2021oes,Acevedo:2021wiq,Dreyer:2021aqd,Duerr:2019dmv,Duerr:2020muu,Filimonova:2019tuy,Chen:2020bok,Dey:2020juy,Bertholet:2021hjl}.
Experimental LLP searches conducted at BaBar and Belle appear in Refs.~\cite{Belle:2013ytx, BaBar:2015jvu}.
We note that Ref.~\cite{Heeck:2017xmg} proposed for the first time to probe LFV with DVs, considering both muon and $\tau$ decays, although no detailed Monte-Carlo (MC) simulation was performed.

The organization of this paper is as follows.
In Sec.~\ref{sec:model} we introduce the effective model under consideration, give the relevant analytic formulas for $\tau$ and $X$ decay rates, and list the two benchmark scenarios we focus on for numerical studies.
We summarize the current constraints on the ALP couplings to the charged leptons in Sec.~\ref{sec:constraints}. In Sec.~\ref{sec:simulation} we briefly describe the Belle II experiment and detail the simulation procedures that we use to study the sensitivity of prompt and displaced $\tau\to Xl$ searches on the ALP couplings.
The numerical results are shown and discussed in Sec.~\ref{sec:results}, and we conclude in Sec.~\ref{sec:conclusions} with a summary and an outlook.

%% file: tex/model.tex

\section{Model}\label{sec:model}

We consider a phenomenological model in the framework of effective field theory of dimension-five for a pseudo-Goldstone boson $X$, corresponding to the ALP considered in this work. The $X$ couples at tree level (only) to the SM charged
leptons $e,\mu$, or $\tau$, with the following Lagrangian~\cite{Feng:1997tn}
\begin{eqnarray}
\mathcal{L}_{X} &=& \frac{\partial_{\mu}X}{\Lambda} \bar{l}_{\alpha} G_{\alpha \beta} \gamma^{\mu} (1+\gamma^{5}) l_{\beta} \nonumber \\
&=& -i\frac{X}{\Lambda}	\bar{l}_{\alpha} G_{\alpha \beta} ((m_{\alpha}-m_{\beta})+(m_{\alpha}+m_{\beta})\gamma^{5}) l_{\beta},\label{eq:Lagrangian}
\end{eqnarray}
where $\Lambda$ labels the effective scale below which the model is applicable, $G$ denotes a symmetric matrix of real couplings ($G_{\alpha\beta} = G_{\beta\alpha}$), and $l_{\alpha,\beta}=e,\mu,\tau$ with mass $m_{\alpha,\beta}$.
To derive the second line of Eq.~\eqref{eq:Lagrangian}, we apply integration by parts and the equations of motion.
For simplicity  we have assumed that the scalar and pseudoscalar couplings are of equal strength for the same lepton flavor combinations.

In order to simplify the notation, we define the following coupling matrix,
\begin{eqnarray}
g_{\alpha \beta} \equiv G_{\alpha \beta}/\Lambda.\label{eq:gab_definition}
\end{eqnarray}
We can thus express the tree-level decay width of one charged lepton $l_\alpha$ into a lighter lepton $l_\beta$ plus $X$ with the formula
\begin{eqnarray}
\Gamma(l_{\alpha} \to l_{\beta} X) &=& \frac{m^{3}_{\alpha}}{8\pi} \sqrt{\left(1-\left(\frac{m_{\beta}+m_{X}}{m_{\alpha}}\right)^{2}\right)\left(1-\left(\frac{m_{\beta}-m_{X}}{m_{\alpha}}\right)^{2}\right)}\nonumber\\
&\times& g^{2}_{\alpha \beta}\left[\left(1-\frac{m^{2}_{\beta}}{m^{2}_{\alpha}}\right)^{2}-\frac{m^{2}_{X}}{m^{2}_{\alpha}}\left(1+\frac{m^{2}_{\beta}}{m^{2}_{\alpha}}\right)\right]. \label{eq:taudecaywidth}
\end{eqnarray}
In the limit $m_{\beta,X} \ll m_{\alpha}$, Eq.~\eqref{eq:taudecaywidth} is reduced to the simple form $m^{3}_{\alpha}g^{2}_{\alpha \beta}/8\pi$.
Our numerical calculations are performed with the full form given in Eq.~\eqref{eq:taudecaywidth}.

The two-body decay widths of $X$ into a pair of charged leptons are given by
\begin{eqnarray}
\Gamma(X \to l_{\beta}^-  l_{\beta}^+) &=& \frac{m_X\, m^2_{\beta}}{2\pi} g^2_{\beta \beta} \sqrt{1-\frac{4m^2_\beta}{m^2_X}},\label{eq:decaywidthX2ll}\\
\Gamma(X\to l_{\alpha}^-  l_{\beta}^+ \text{ or } l_{\alpha}^+  l_{\beta}^-)&\simeq&\frac{m_X\, m^2_{\alpha}}{2\pi} g_{\alpha\beta}^2\Big( 1 - \frac{m_\alpha^2}{m_X^2} \Big)^2,\label{eq:decaywidthX2ll2}
\end{eqnarray}
where the two equations are for decays into same-flavor and different-flavor charged leptons, and the second equation is obtained in the limit $m_\beta \ll m_\alpha$.

We note that the flavor-diagonal couplings $g_{\alpha\alpha}$ can induce $X$ decays into a pair of photons, as well as those into a pair of neutrinos, both at one-loop order.
Compared to the di-photon channel, the di-neutrino decay is highly suppressed (see the relevant discussions in Ref.~\cite{Bauer:2017ris}) and is hence ignored in this work.
The corresponding decay width can be calculated with: \cite{Bauer:2017ris}
\begin{eqnarray}
\Gamma(X \to \gamma\gamma) = 4\pi \alpha^2 m_X^3 |g^{\text{eff}}_{\gamma\gamma}|^2,\label{eq:decaywidthX2gammagamma}
\end{eqnarray}
where $g_{\gamma\gamma}^{\text{eff}}$ is an effective coupling that can be derived from the ALP-lepton couplings:
\begin{eqnarray}
g^{\text{eff}}_{\gamma \gamma} = \frac{1}{8 \pi^2}\sum_{\alpha=e,\mu,\tau}g_{\alpha\alpha} \, B_1(4 m_\alpha^2/m_X^2),\label{eq:decaywidthX2gg}
\end{eqnarray}
where the loop function $B_1$ is defined as follows:
\begin{eqnarray}
  B_{1}(t) &=& 1-t f^{2}(t), \text{ with } f(t) = \begin{cases}
	\arcsin(\frac{1}{\sqrt{t}}),& \text{if } t \geq 1 \\
	\frac{i}{2}\ln(\frac{1+\sqrt{1-t}}{1-\sqrt{1-t}})+\frac{\pi}{2},& \text{if } t < 1 \end{cases}.
\end{eqnarray}
Note that both the tree-level couplings $g_{\alpha\alpha}$ and the effective coupling $g_{\gamma\gamma}^{\text{eff}}$ are of mass dimension $-1$, \textit{cf.} Eq.~\eqref{eq:gab_definition}.
With Eqs.~\eqref{eq:decaywidthX2ll},~\eqref{eq:decaywidthX2ll2}, and~\eqref{eq:decaywidthX2gg}, we can compute the total decay width and, hence, the proper lifetime of the ALP as a function of its mass and the couplings to the charged leptons.

In this paper, we concentrate on the scenarios where we switch on two couplings: one dominating the production and another governing the decay of the ALP.
Specifically, the ALP is produced in the decay $\tau\to l_\alpha X$ via the production couplings $g_{\tau \alpha}$ ($\alpha=e,\mu$), and  undergoes the decay $X\to e^-e^+$ or $\mu^+\mu^-$ via the decay couplings $g_{\alpha\alpha}$.
For simplicity, we do not consider $X$ decays via the off-diagonal coupling $g_{e\mu}$,\footnote{For a study on this coupling and LFV at Belle II, we refer the reader to Ref.~\cite{Endo:2020mev}.} but will offer brief comments when we discuss the numerical results in Sec.~\ref{sec:results}.
Moreover, the $g_{\tau\tau}$ coupling is also switched off, so we do not consider ALP production by final-state radiation.

We should point out that in principle, the production coupling  $g_{\tau \alpha}$ alone can also induce (lepton/quark-level) four-body $X$ decays via an off-shell $\tau$, as well as, sub-dominantly, the two-body decays $X\to \nu_\tau\bar{\nu}_\alpha$ (or their CP conjugate) via a triangular loop  with a $W$-boson.
However, these contributions are negligible compared to those from the tree-level decays considered here, unless $g_{\tau \alpha}\gg  g_{\alpha\alpha}$.
In the case that the diagonal couplings are indeed negligible, the ALP becomes very long-lived, requiring  a missing-energy search strategy ~\cite{ARGUS:1995bjh,Tenchini:2020njf} that is not the topic of this paper.
This particular caveat will not affect our numerical results, and so the highly suppressed $X$ decays via $g_{\tau \alpha}$ couplings are not included in the numerical estimates.
For $m_X > m_{\tau}$, limits on $g_{\tau \alpha}$ can be derived from $e^- e^+ \to l^\pm l^\pm \tau^\mp \tau^\mp$ and $e^- e^+ \to l^\pm \tau^\mp +\text{ missing}$~\cite{Iguro:2020rby}.

\subsection{Benchmark Scenarios}\label{subsec:benchmark_scenarios}
In our numerical studies, we  focus on two representative benchmark scenarios.
In Scenario 1, $g_{\tau e}$ and $g_{ee}$ are switched on, and the other couplings are set to zero.
As a result, $g_{\tau e}$ induces the decay $\tau\to Xe$ and $g_{ee}$ leads to $X\to e^-e^+$ and $\gamma\gamma$.
In  Scenario 2, only $g_{\tau \mu}$ and $g_{\mu\mu}$ are non-zero, so that the ALP is produced via $\tau\to X\mu$ and decays to $\mu^-\mu^+$ and $\gamma\gamma$.
We focus on the di-lepton final state and do not consider the di-photon decay mode as a signature of interest in this work.
We summarize these two scenarios in Table~\ref{tab:scenarios}.
\begin{table}[htb]
	\centering
	\begin{tabular}{c|c|c}
		& Scenario 1  		&  Scenario 2 \\
		\hline
		$g_{\tau \alpha}$	& $g_{\tau e}$ 	& $g_{\tau\mu}$ \\
		tau decays	 	& $\tau\to Xe$ &  $\tau \to X\mu$\\
		\hline
		$g_{\beta \beta}$	& $g_{ee}$	    & $g_{\mu\mu}$  \\
		   $X$ decays		& $X\to e^- e^+(\text{sig.})/\gamma\gamma$  & $X\to \mu^-  \mu^+(\text{sig.})/\gamma\gamma$
	\end{tabular}
	\caption{Summary of the basic features of the two benchmark scenarios studied in this work, including the couplings switched on and the corresponding induced decays of $\tau$ and $X$.
	``Sig.'' indicates the signature studied.
}
	\label{tab:scenarios}
\end{table}

\begin{figure}[htb]
	\centering
	\includegraphics[width=0.5\textwidth]{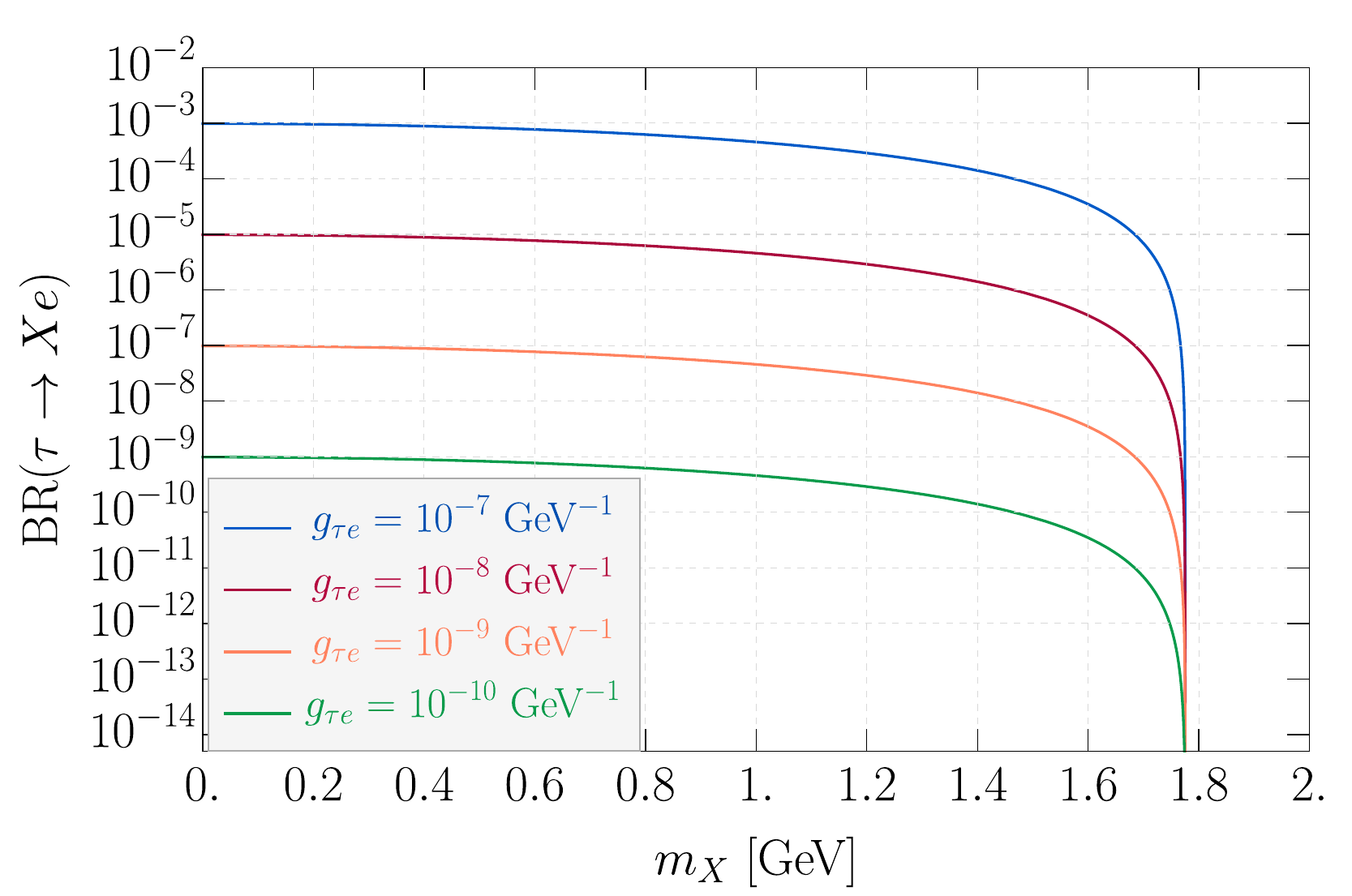}\\
	\includegraphics[width=0.49\textwidth]{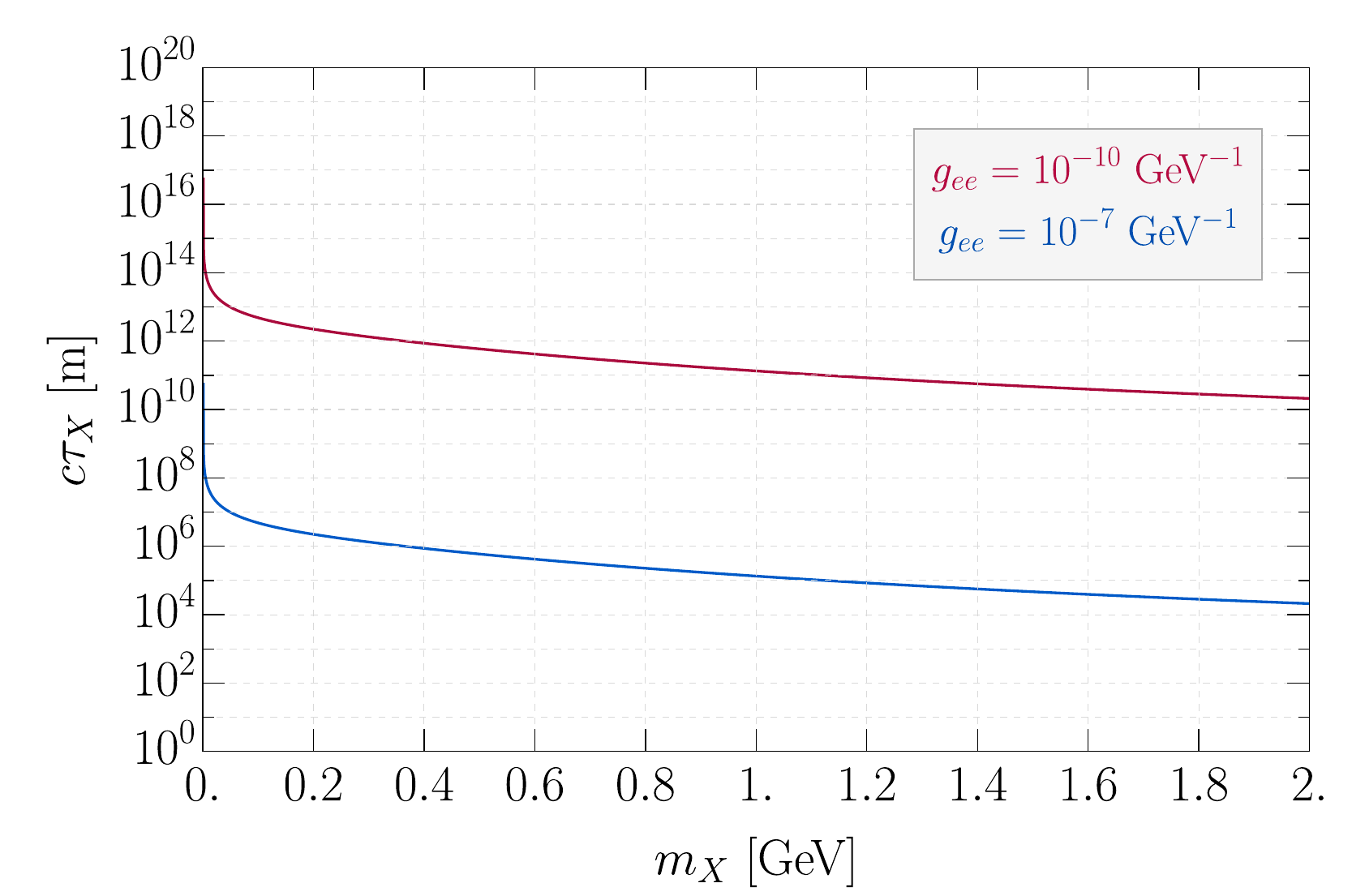}
	\includegraphics[width=0.49\textwidth]{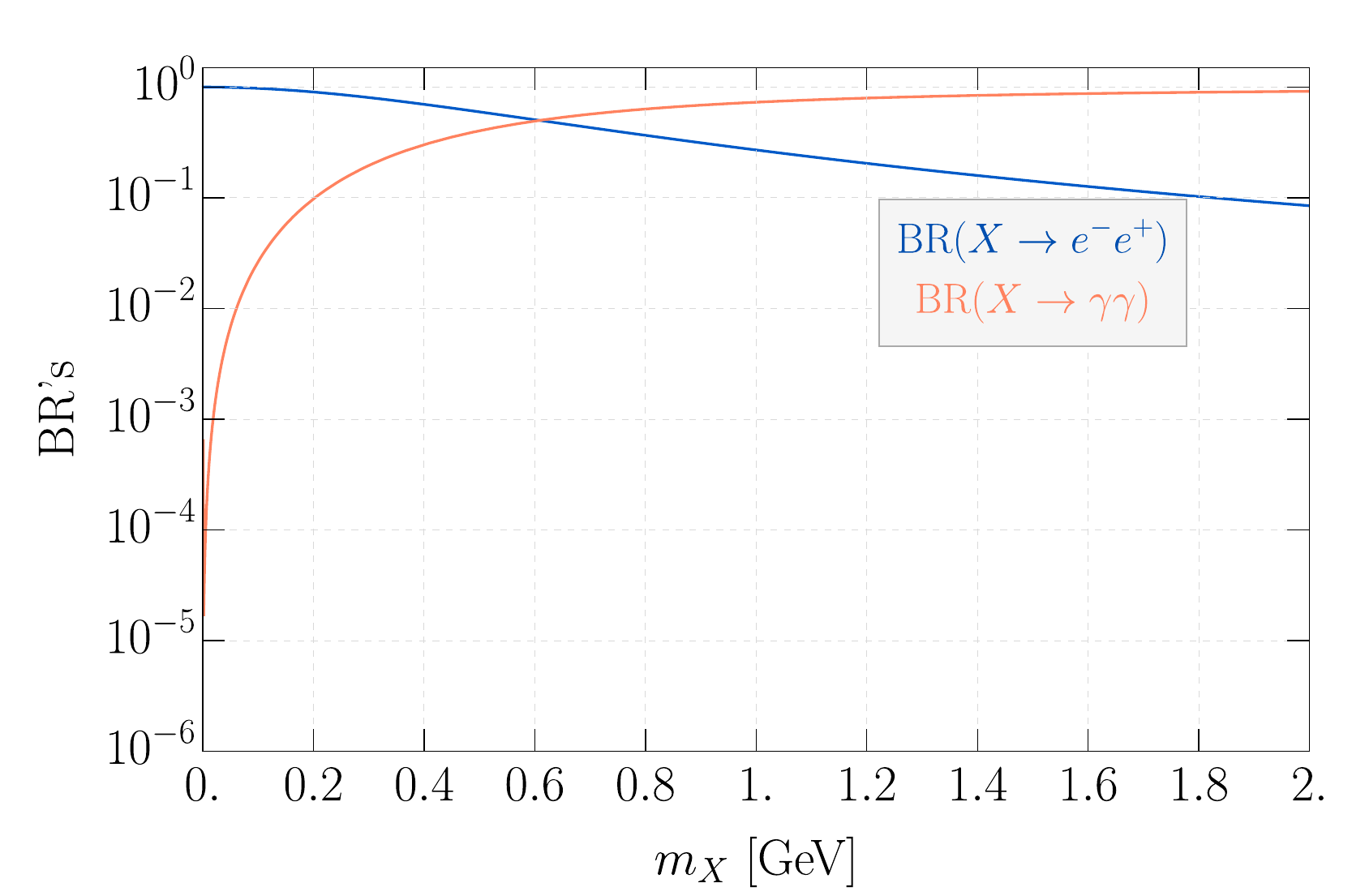}
	\caption{Scenario 1. \textit{Top}: branching ratio of $\tau\to Xe$ vs.~$m_X$ for $g_{\tau e}=10^{-7},10^{-8},10^{-9},10^{-10}$ GeV$^{-1}$.
	  \textit{Bottom left}: ALP proper flight distance $c\tau_X$ vs.~$m_X$ for $g_{ee}=10^{-10},10^{-7}$ GeV$^{-1}$.
	   \textit{Bottom right}:  branching ratios of $X\to e^- e^+$ or
           $\gamma\gamma$, as a function of $m_X$.	
	}
	\label{fig:BRs_scenario1}
\end{figure}
\begin{figure}[t]
	\centering
	\includegraphics[width=0.49\textwidth]{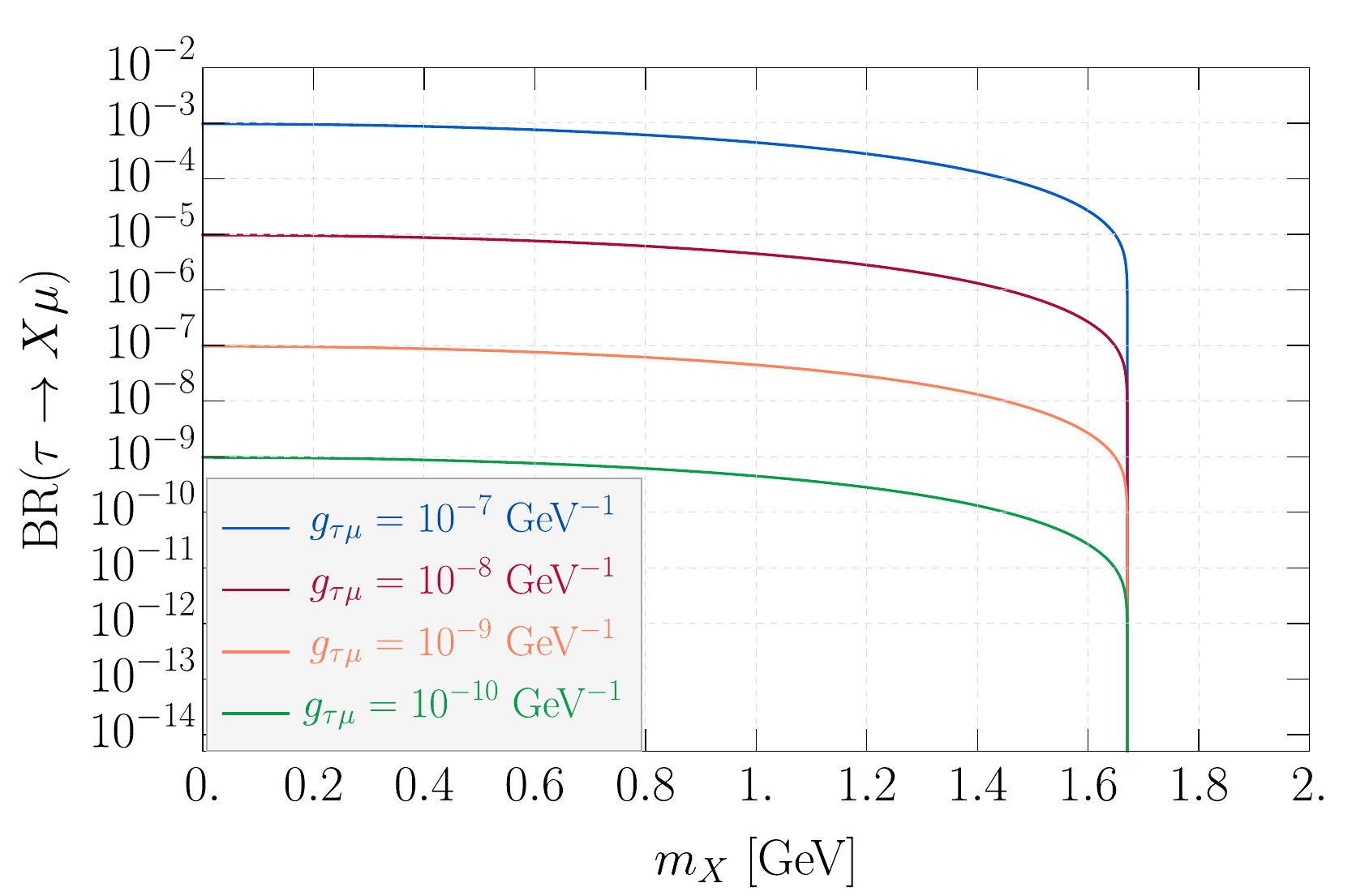}\\
	\includegraphics[width=0.49\textwidth]{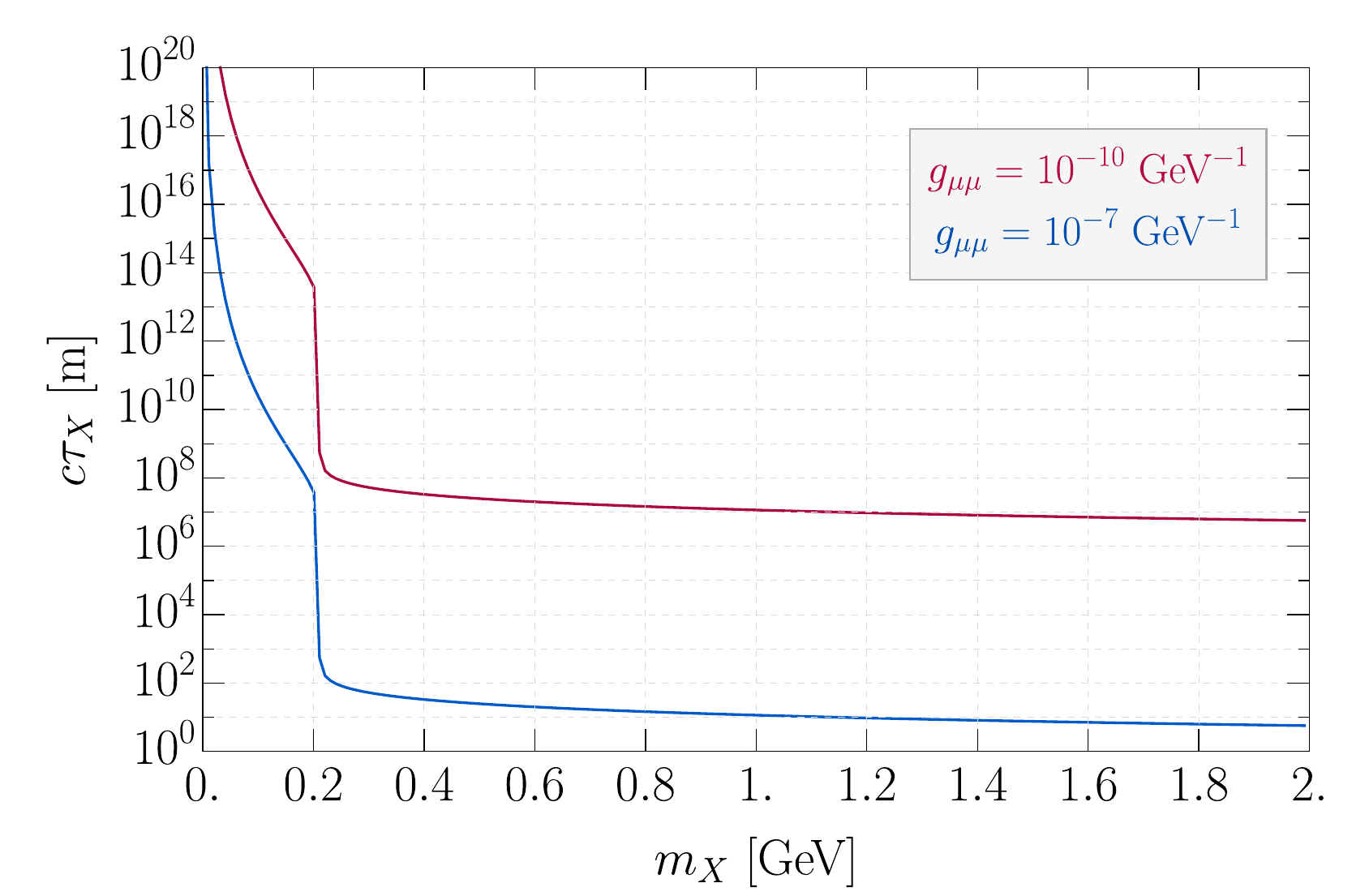}
	\includegraphics[width=0.49\textwidth]{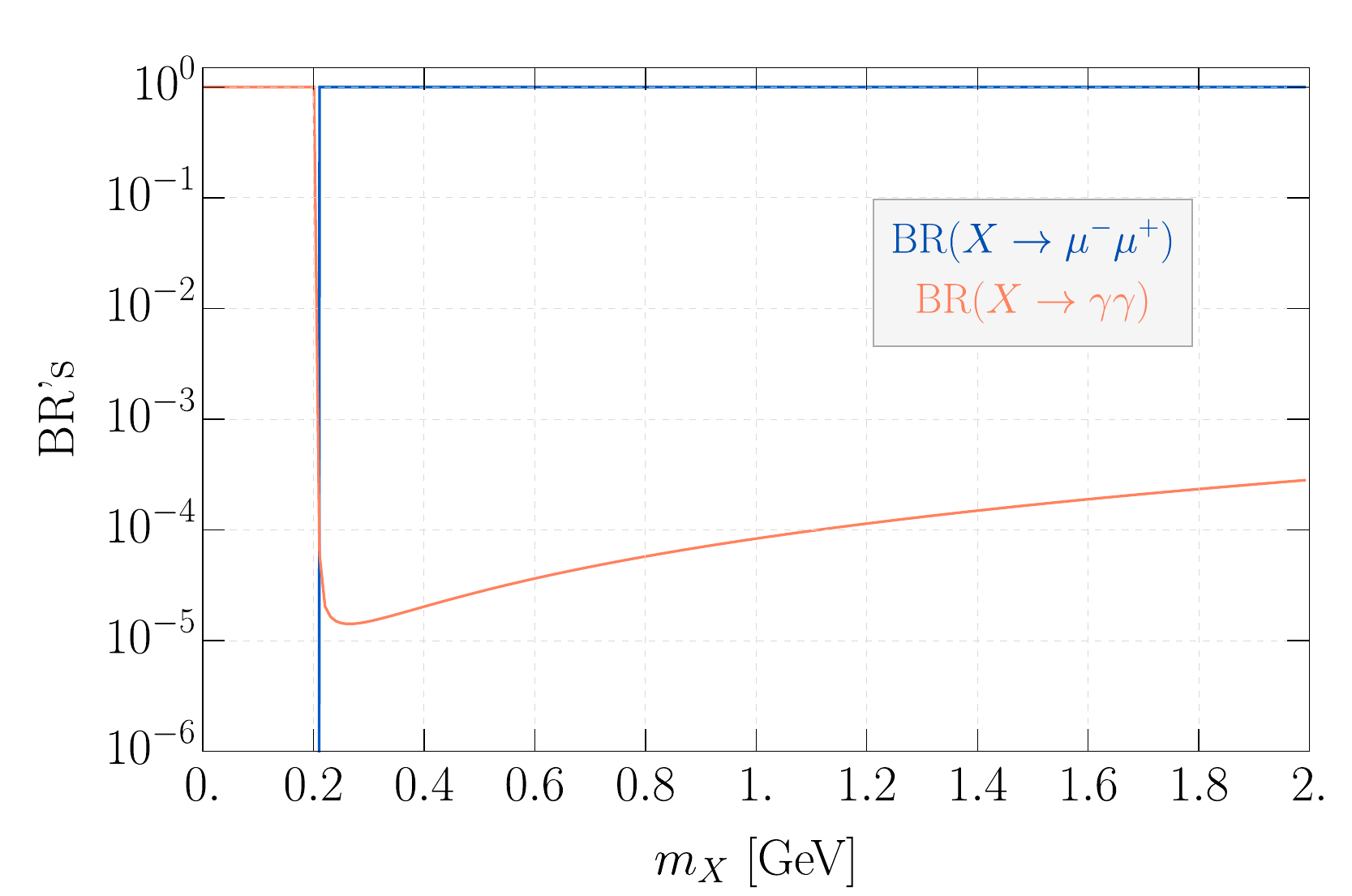}
	\caption{Scenario 2. The same format as in Fig.~\ref{fig:BRs_scenario1}, replacing $e$ by $\mu$. 
	}
	\label{fig:BRs_scenario2}
\end{figure}
In Fig.~\ref{fig:BRs_scenario1} and Fig.~\ref{fig:BRs_scenario2} we present plots of the branching fraction BR$(\tau\to Xl_\alpha)$, the ALP proper flight distance $c\tau_X$, and its decay branching fractions as functions of $m_X$ for the two benchmark scenarios.
The upper plots of Fig.~\ref{fig:BRs_scenario1} and Fig.~\ref{fig:BRs_scenario2} show the variation of BR$(\tau\to X l_\alpha)$ with respect to $m_X$, for several benchmark values
of $g_{\tau \alpha}$. 
Intriguingly, the $X$ branching fraction plots show that BR$(X\to \gamma  \gamma)$ exceeds BR$(X\to e^-  e^+)$ for $m_X \gtrsim 0.6$ GeV in Scenario 1, despite the loop suppression in $(X\to \gamma \gamma)$.
This observation can be understood from the fact that $\Gamma(X\to l_\alpha^-  l_\alpha^+) \propto m_X m_\alpha^2$, while  $\Gamma(X\to \gamma \gamma) \propto m_X^3$,  \textit{cf}.~Eqs.~\eqref{eq:decaywidthX2ll2} and~\eqref{eq:decaywidthX2gg}.
Therefore, given the tiny electron mass, $\Gamma(X\to e^-  e^+)$ can become subdominant for sufficiently large $m_X$.

Besides the two benchmark scenarios introduced above, there are, of course, additional scenarios with other possible lepton flavor combinations.
In Sec.~\ref{subsec:other_scenarios}, we will qualitatively comment on their numerical results instead of performing detailed studies.

%% file: tex/constraints.tex

\section{Constraints}\label{sec:constraints}

In this section, we discuss various experimental and astrophysical results from which we derive constraints on the model parameters related to Scenarios 1 and 2 introduced in Sec.~\ref{sec:model}.
The experimental results come from LEP, $B$-factories, and muon fixed-target experiments, while the astrophysical bounds stem mainly from the supernova 1987A (SN1987A).
We also refer the reader to Ref.~\cite{Cornella:2019uxs} for a comprehensive study on probes  of ALPs with LFV.

\subsection{LEP}

The LEP-2 collider at CERN provided samples of $e^- e^+\to \tau^-\tau^+$ events
at center-of-mass energies $\sqrt{s}$ ranging from 189 to 209~GeV.
The inclusive cross section of $e^- e^+\to \tau^-\tau^+$ at $\sqrt{s}=207$ GeV was measured to be $\sigma_{\tau\tau}^{\text{LEP}}=6.05\pm0.39 \, {\rm (stat)}\,
\pm0.09 \,{\rm (syst)}$ pb with 402 observed events at the ALEPH experiment~\cite{ALEPH:2006jhv}, in good agreement with the SM predictions.
If the $g_{\tau e}$ coupling is non-zero, it can induce the same process via a $t$-channel ALP exchange, modifying the theoretical computation of the inclusive cross section.
Thus, we use the experimental measurement to place a bound on the coupling $g_{\tau e}$.

We implement the ALP model (see Lagrangian in Eq.~\eqref{eq:Lagrangian}) using FeynRules~\cite{Christensen:2008py,Alloul:2013bka} to generate an UFO (Universal FeynRules Output)~\cite{Degrande:2011ua} model file, with which we simulate $e^- e^+\to \tau^-\tau^+$ events and compute the corresponding cross sections for given $m_X$ and $g_{\tau e}$ with MadGraph5 2.7.3~\cite{Alwall:2014hca}.
By scanning over a 2D grid of $(m_X,g_{\tau e})$, we exclude the points where the computed cross section lies outside the two-sided range around the central value of the ALEPH measurement: $[\sigma_{\tau\tau}^{\text{LEP}}-2\Delta\sigma, \ \sigma_{\tau\tau}^{\text{LEP}}+2\Delta\sigma]$, where $\Delta\sigma \simeq \sqrt{0.39^2+0.09^2}\text{ pb}=0.40\text{ pb}$ is the total experimental uncertainty on $\sigma_{\tau\tau}^{\text{LEP}}$. The theoretical uncertainty is negligible and hence neglected.
For $m_X\lesssim 2$ GeV the limit is roughly a constant at $g_{\tau e}< 0.078$ GeV$^{-1}$, because of the relatively large center-of-mass energy.
This is a very weak constraint relative to others studied here.

We also note that if one switches on the coupling $g_{\tau\tau}$ in Scenario 1 together with $g_{ee}$, it can induce $\tau$-pair production via an $s$-channel $X$ in $e^- e^+$ collisions, thus allowing to probe the product of these two couplings.
Such a scenario is beyond the scope of this work.
Moreover, the $g_{ee}$ coupling itself affects M\o ller and Bhabha scattering. A precise measurement of the parity violation asymmetry in M\o ller scattering has been
performed in a fixed target experiment~\cite{SLACE158:2005uay}.
Nevertheless, the axion model under consideration only has the pseudoscalar coupling to electron after taking the derivative of the field $X$ (see the second line of Eq.~(\ref{eq:Lagrangian}) ), so it does not contribute to parity violation.

A limit on the ALP interactions using Bhabha scattering excludes values of $\Gamma(X)[\text{BR}(X\to e^- e^+)]^2$ above $1.3\cdot 10^{-3}$ eV for $m_X$ between 1.75 and 1.88 MeV~\cite{Hallin:1992sj}. This can be translated to a weak bound of $g_{ee}\gtrsim 0.15$ GeV$^{-1}$ for $m_X=1.8$ MeV in our model, \textit{cf}.~Eqs.~\eqref{eq:decaywidthX2ll} and~\eqref{eq:decaywidthX2gammagamma}.

The coupling $g_{\tau\mu}$ and the coupling product  $
g_{\mu\mu} g_{\tau\tau}$ could in principle be studied at future high-energy muon
colliders~\cite{Haghighat:2021djz,Delahaye:2019omf,Boscolo:2018ytm,AlAli:2021let} by measuring the process $\mu^- \mu^+ \to \tau^- \tau^+$.

\subsection{Supernova}

If the ALP couples to electrons or muons, it can be produced in the supernova environment and contribute to the cooling rate of the supernova core, thus allowing for deriving bounds on these couplings.
For instance, Ref.~\cite{Bollig:2020xdr} obtained an upper bound on $g_{\mu\mu}$ at $10^{-8.1}$ GeV$^{-1}$ for $m_X<1$ MeV.
Also, Ref.~\cite{Lucente:2021hbp} derived the limits on $g_{ee}$ with SN1987A ranging
from $\sim 2.5\times 10^{-7}$ GeV$^{-1}$ to $\sim 8.8\times 10^{-5}$ GeV$^{-1}$ for $m_X\lesssim 160$ MeV.
\begin{figure}[htb]
	\centering
	\includegraphics[width=0.49\textwidth]{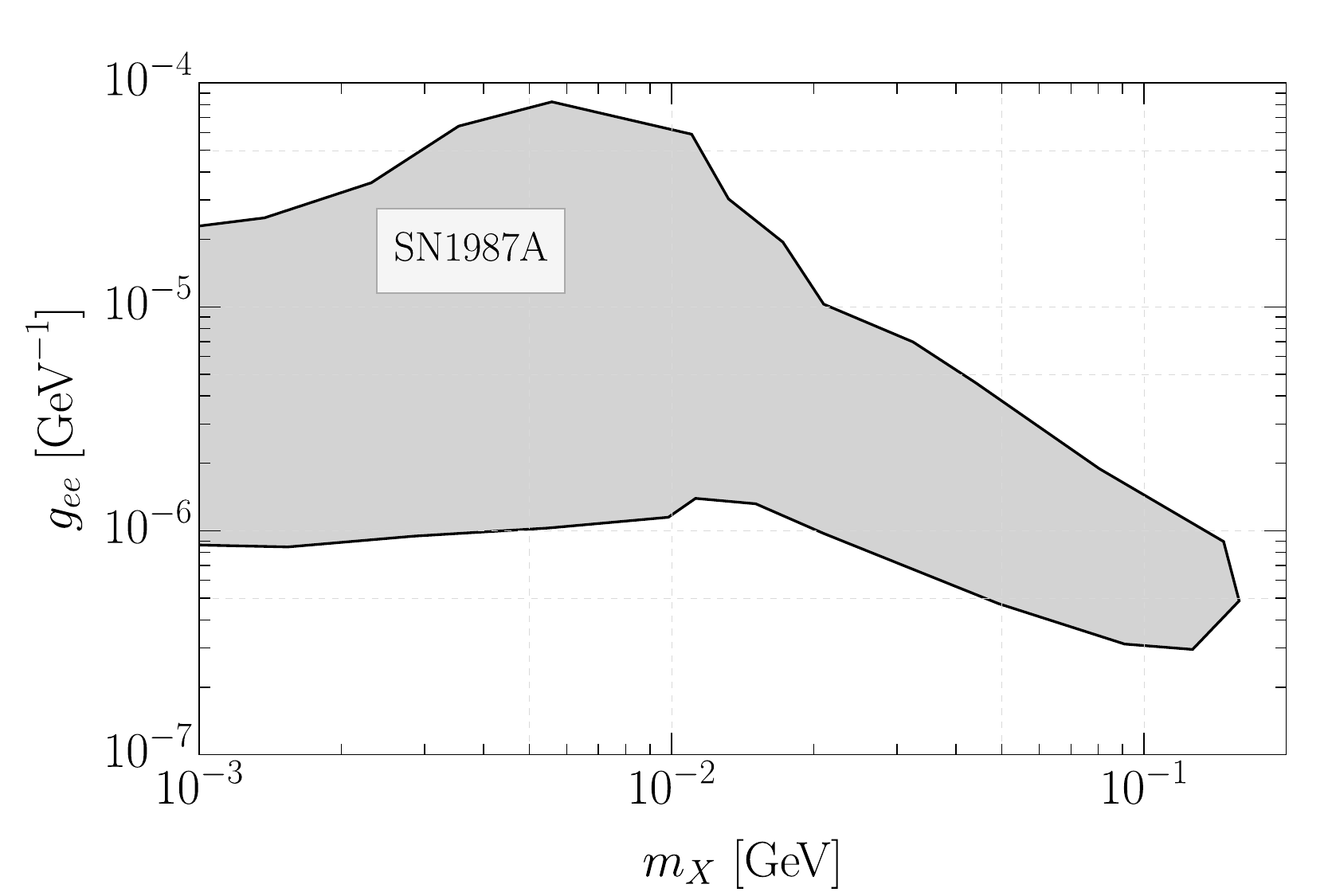}
	\caption{Constraint on $g_{ee}$ with respect to $m_X$, derived in Ref.~\cite{Lucente:2021hbp} from the supernova cooling efficiencies.
	}
	\label{fig:constraints_supernova}
\end{figure}
In our relevant mass range, we reproduce in Fig.~\ref{fig:constraints_supernova} the SN1987A bounds on $g_{ee}$ here\footnote{The coupling `$g_{ae}$' in Ref.~\cite{Lucente:2021hbp} divided by $2m_e$ yields $g_{ee}$ of this paper.}.
The gray region is disfavored by SN1987A cooling bounds.

\subsection{$\tau\to 3l$}
\label{sec:constraints-tau3l}
The tightest bounds on the branching fractions for $\tau$ LFV decays into three charged
leptons were obtained at the Belle experiment~\cite{Hayasaka:2010np}.
These limits, which are all of order $10^{-8}$, are reproduced in Table~\ref{tab:tau23l_upper_bounds}.
We discuss the resulting bounds on $g_{\tau \alpha}\cdot g_{\alpha\alpha}$ ($\alpha=e,\mu$) in Sections~\ref{sec:simulation} and~\ref{sec:results}.
\begin{table}[htb]
	\centering
	\begin{tabular}{c|c}
		Decay modes	& Upper bounds on BR $[10^{-8}]$  	\\
		\hline
		$\tau^- \to e^- e^+ e^-$	 		& $2.7$ \\
		$\tau^- \to \mu^- \mu^+ \mu^-$		& $2.1$ \\
		$\tau^- \to e^- \mu^+ \mu^-$			& $2.7$ \\
		$\tau^- \to \mu^- e^+ e^-$			& $1.8$ \\
		$\tau^- \to e^+ \mu^- \mu^-$			& $1.7$ \\
		$\tau^- \to \mu^+ e^- e^-$			& $1.5$ \\
	\end{tabular}
	\caption{Upper bounds on the branching fractions of LFV decays of the $\tau$ into three charged
          leptons, reproduced from Ref.~\cite{Hayasaka:2010np}.}
	\label{tab:tau23l_upper_bounds}
\end{table}

\subsection{Lepton universality in $\tau \to l \nu \bar{\nu}$}\label{subsec:lepton_universality}
When the ALP lifetime is long enough that the ALP rarely decays before traversing the whole detector, the decay $\tau\to l_\alpha X$ yields a signature similar to that of $\tau\to l\nu\bar{\nu}$, namely $\tau\to l+\text{missing}$.
It is thus possible to place bounds on the off-diagonal coupling $g_{\tau \alpha}$ for long-lived ALPs, in the scenario $g_{\tau \alpha} \gg g_{\tau \beta}$.
For the coupling $g_{\tau \mu}$, inspired by Refs.~\cite{Foldenauer:2016rpi,Altmannshofer:2016brv}, we focus on the ratio of rates
\begin{eqnarray}
R_{\mu e}=\frac{\Gamma_{\tau\to\mu\nu\bar{\nu}}}{\Gamma_{\tau\to e \nu\bar{\nu}}}\,,
\end{eqnarray} 
which has been computed accurately in the SM~\cite{Pich:2013lsa}:
\begin{eqnarray}
	R^{\text{SM}}_{\mu e}= 0.972559 \pm 0.000005.
\end{eqnarray}
The BaBar collaboration measured this ratio to be~\cite{BaBar:2009lyd}
\begin{eqnarray}
R_{\mu e}^{\text{BaBar}} = 0.9796 \pm 0.0039.
\end{eqnarray}
We define the discrepancy between $R^{\text{SM}}_{\mu e}$ and $R_{\mu e}^{\text{BaBar}}$,
\begin{eqnarray}
\Delta R_{\mu e} \equiv R_{\mu e}^{\text{BaBar}}/R^{\text{SM}}_{\mu e}-1 = 0.0072\pm 0.0040. \label{eq:DeltaRmue}
\end{eqnarray}
We also define the theoretical value of this ratio in the presence of the ALP,         
	\begin{eqnarray}
	  R^{\text{SM}+X} = R^{\text{SM}}_{\mu e} + \Gamma(\tau \to X\mu)/\Gamma_{\tau\to e \nu\bar{\nu}}^{\text{SM}}
          \;,
	\end{eqnarray}
	where we use the experimental result $\Gamma^{\text{exp}}_{\tau\to e\nu\bar{\nu}}=4.04\times 10^{-13}$
        GeV~\cite{ParticleDataGroup:2020ssz} for $\Gamma^{\text{SM}}_{\tau\to e\nu\bar{\nu}}$,
and $\Gamma(\tau\to X\mu)$ is computed with Eq.~\eqref{eq:taudecaywidth}.
We determine the 95\% confidence-level (C.L.) bounds on $g_{\tau \mu}$ vs. $m_X$ from the region that satisfies the following condition:
\begin{eqnarray}
R_{\mu e}^{\text{SM}+X}/R^{\text{SM}}_{\mu e}-1<0.0072+2 \times 0.0040,
\end{eqnarray}
where 0.0072 is the central value of $\Delta R_{\mu e}$ given in Eq.~\eqref{eq:DeltaRmue} and $0.0040$ is its uncertainty.

To derive bounds on $g_{\tau e}$ we follow a similar procedure with the ratios $R_{e \mu}=1/R_{\mu e}$.
In this case, since $R_{e\mu}^{\text{SM}}$ is now larger than $R_{e\mu}^{\text{BaBar}}$, we have
	\begin{eqnarray}
	\Delta R_{e\mu}\equiv R_{e\mu}^{\text{BaBar}}/R_{e\mu}^{\text{SM}} - 1 =-0.0072\pm 0.0040.
	\end{eqnarray}
The negative value of $\Delta R_{e\mu}$ results in a particularly stringent bound on BR$(\tau\to X e)$, after requiring $R_{e \mu }^{\text{SM}+X}/R^{\text{SM}}_{e \mu}-1<-0.0072+2 \times 0.0040$.
\begin{figure}[htb]
	\centering
	\includegraphics[width=0.49\textwidth]{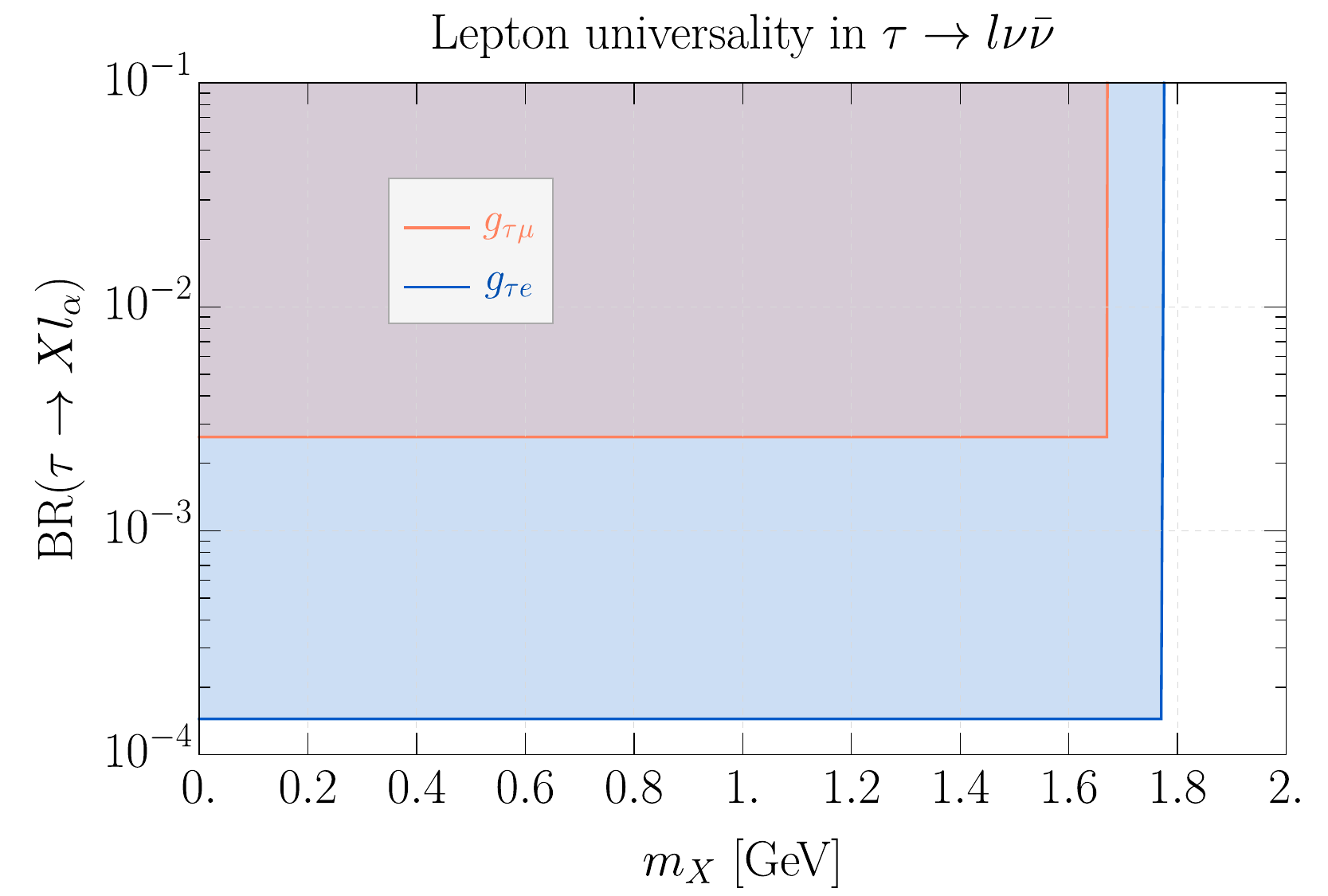}
	\includegraphics[width=0.49\textwidth]{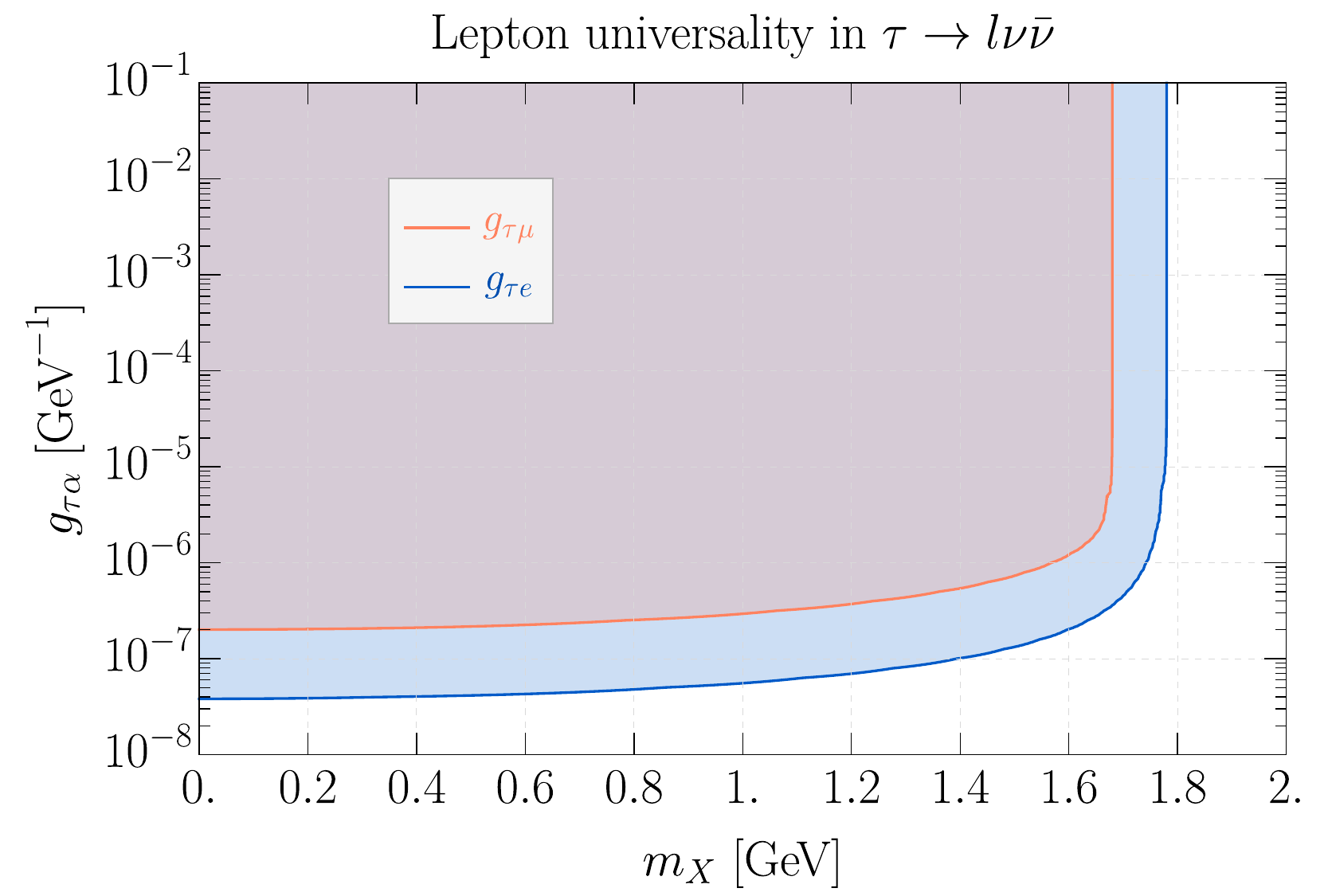}
	\caption{ Constraints on BR$(\tau\to X l_\alpha)$ (left) and on $g_{\tau \alpha}$ (right) as a function of $m_X$, derived from the upper bounds on lepton-universality violation in $\tau\to l+\text{missing}$, for Scenarios 1 ($g_{\tau e}\ne 0$) and 2 ($g_{\tau \mu}\ne 0$).
	  These limits are valid for long decay lengths of the $X$ (small diagonal couplings $g_{ee}$ or $g_{\mu\mu}$), so that it rarely decays inside the detector.
	}
	\label{fig:constraints_tau2lmissing}
\end{figure}
The resulting constraints are shown in Fig.~\ref{fig:constraints_tau2lmissing}.
For the kinematically allowed mass range, the constraints on BR$(\tau\to X l_\alpha)$ are about $1.4\times 10^{-4}$ for Scenario 1 and $2.6\times 10^{-3}$ for the second scenario.
The corresponding upper limits on $g_{\tau \alpha}$ are as strong as about 
 $4\times 10^{-8}$ GeV$^{-1}$ and $2\times 10^{-7}$ GeV$^{-1}$ for the two scenarios, respectively, at the low mass regime.
After the limits described in Sec.~\ref{sec:constraints-tau3l}, these are the tightest existing limits.
We note that they are valid for small diagonal couplings, so that the ALP decays inside the detector only rarely.

It should be noted that the lepton-universality constraint will improve after the analysis of Ref.~\cite{BaBar:2009lyd} is performed with the large Belle~II data set.
However, the results of Ref.~\cite{BaBar:2009lyd} are already limited by the systematics uncertainties, the improvement is expected to be modest.

We notice that Ref.~\cite{Bryman:2021rtr} has also recast the BaBar search results to the upper bounds on BR$(\tau \to X l_{\alpha})$. 
In contrast to our treatment, which takes into account the experimental measurements of the ratios $R_{e\mu}$ and $R_{\mu e}$ of $\Gamma_{\tau\to \mu\nu\bar{\nu}}$ and $\Gamma_{\tau\to e \nu\bar{\nu}}$, the authors of Ref.~\cite{Bryman:2021rtr} worked on the partial decay widths of the two decay channels separately. 
As a result, our bound on BR$(\tau\to X e)$ is stronger than theirs by about one order of magnitude, while both works obtained similar constraints on BR$(\tau \to X \mu)$.

\subsection{$\tau\to l\gamma$}

The ALP couplings to the charged leptons can also lead to the decays $\tau \to l_\alpha \gamma$ via either one or two LFV couplings.
In the case of two LFV couplings, the processes must involve the coupling $g_{\mu e}$, which we set to zero throughout the paper.
  Therefore, we focus on the combinations of one LFV coupling  and one diagonal coupling.
The relevant coupling combinations are $(g_{\tau\tau},g_{\tau e})$ and $(g_{\tau e},g_{e e})$ for Scenario 1, and $(g_{\tau\tau}, g_{\tau \mu})$ and $(g_{\tau \mu},g_{\mu\mu})$ for Scenario 2.
Since we take $g_{\tau \tau} =0$ in this paper, we derive only bounds on the coupling
products $g_{\tau e}\cdot g_{ee}$ and $g_{\tau\mu}\cdot g_{\mu\mu}$ from the experimental upper bounds
BR$(\tau\to e\gamma)<3.3\times 10^{-8}$ and BR$(\tau\to \mu\gamma)<4.4\times 10^{-8}$~\cite{BaBar:2009hkt}, respectively.

Under these conditions, the ALP contributions to these decays consist of two categories: two-loop diagrams with an effective $g_{\gamma\gamma}^{\text{eff}}$ coupling, and one-loop diagrams without this coupling.
The former turns out to be dominant because of the logarithmic dependence on the ultraviolet cutoff.
The decay width of $\tau\to l_\alpha \gamma $ is computed with the following analytic expression~\cite{Bauer:2019gfk,Cornella:2019uxs}:
\begin{eqnarray}
\Gamma(\tau\to l_\alpha \gamma)=\frac{m_\tau e^2}{8\pi}\Big( |\mathcal{F}_2^\alpha(0)|^2 +|\mathcal{G}_2^\alpha(0)|^2\Big),\text{ for }m_\tau \gg m_e, m_\mu,
\end{eqnarray}
where $\alpha=e,\mu$, and $\mathcal{F}_2^\alpha(0)$ and $\mathcal{G}_2^\alpha(0)$ are dimensionless form factors
 consisting of two parts that are naively linear and quadratic in the ALP-lepton couplings:
\begin{eqnarray}
\mathcal{F}_2^\alpha(0)&=&\mathcal{F}_2^\alpha(0)_{\text{lin.}}+\mathcal{F}_2^\alpha(0)_{\text{quad.}},\nonumber\\
\mathcal{G}_2^\alpha(0)&=&\mathcal{G}_2^\alpha(0)_{\text{lin.}}+\mathcal{G}_2^\alpha(0)_{\text{quad.}},
\end{eqnarray}
where 
\begin{eqnarray}
 \mathcal{F}_2^{\alpha}(0)_{\text{lin.}} &=& -\frac{e^2 m_\tau^2}{8\pi^2}g_{\tau \alpha} g_{\gamma\gamma}^{\text{eff}} g_\gamma(x_\tau),\\
 \mathcal{G}_2^{\alpha}(0)_{\text{lin.}} &=& -\frac{e^2 m_\tau^2}{8\pi^2}g_{\tau \alpha} g_{\gamma\gamma}^{\text{eff}} g_\gamma(x_\tau),\\
 \mathcal{F}_2^{\alpha}(0)_{\text{quad.}} &=& - \frac{m_{\tau}}{16\pi^2}\,g_{\tau \alpha}\,g_{\alpha\alpha}\,m_{\alpha}\,g_2(x_\tau), \nonumber\\ \mathcal{G}_2^{\alpha}(0)_{\text{quad.}} &=& + \frac{m_{\tau}}{16\pi^2}\,g_{\tau \alpha}\,g_{\alpha\alpha}\,m_{\alpha}\,g_2(x_\tau),	
\end{eqnarray}
where $x_\tau=m_X^2/m_\tau^2-i\epsilon$ with $\epsilon \to 0^+$ and
\begin{eqnarray}
	g_\gamma(x)&=&2\text{log}\frac{\Lambda^2}{m_X^2}-\frac{\text{log}x}{x-1}-(x-1)\text{log}\frac{x}{x-1}-2, \label{eq:gammafunction} \\
	g_2(x)&=&1-2x+2(x-1)x\, \text{log}\frac{x}{x-1}.
\end{eqnarray}
We have assumed $g_{e\mu}=g_{\mu e}=g_{\tau \tau} = 0$, and confined ourselves only to the couplings that are turned on in Scenarios 1 and 2 for simplicity.
These couplings, if non-zero,  could also lead to $\tau\to l_\alpha \gamma$ decays with two ALP-lepton couplings.
For full and complete formulas, see Refs.~\cite{Bauer:2019gfk,Cornella:2019uxs}.

The resulting bounds on $g_{\tau e} g_{ee}$ and $g_{\tau \mu} g_{\mu\mu}$ are roughly $10^{-6}$ GeV$^{-2}$ for $m_X$ between 0 and 2 GeV.
Note that we have fixed $\Lambda=1$ TeV for the logarithmic dependency (see Eq.~\eqref{eq:gammafunction}).

\subsection{Leptonic decays of the muon}

Multiple processes of muon decays into electrons or photons can be induced by the ALP couplings.
For instance, the decay $\mu\to e\gamma$~\cite{Petcov:1976ff,Hernandez-Tome:2018fbq}, which is extremely rare in the SM, can be mediated either by $g_{\mu e}$ together with $g_{ee}$ or $g_{\mu \mu}$, or by the two LFV couplings $g_{\tau \mu}$ and $g_{\tau e}$.
The experimental limits on BR($\mu\to e\gamma$)~\cite{MEG:2016leq} place  very strong bounds on these coupling products.
Since in this work we take $g_{\mu e}=0$ and do not switch on $g_{\tau\mu}$ and $g_{\tau e}$ simultaneously, both of these two contributions vanish (see Eqs.~(11-14) of Ref.~\cite{Cornella:2019uxs}).

Similarly, the processes $\mu\to 3e$, $\mu\to e\gamma\gamma$, and $\mu\to e+\text{missing}$ can all be used to derive bounds on the coupling $g_{\mu e}$ (and $g_{ee}$).
However, since we have restricted  ourselves to the case of vanishing $g_{\mu e}$, we do not study these processes in detail.

\subsection{Leptonic $g-2$ anomalies}

One of the most sensitive avenues for testing the SM is the leptonic anomalous magnetic moments.
The long-standing discrepancy in the muon $g-2$ between the SM predictions and experimental
measurements has been an important drive for searching for new physics.
The latest determination of the fine-structure constant $\alpha_{\text{EM}}$ makes the electron anomalous magnetic moment consistent with the SM prediction to within $1.6\sigma$~\cite{Morel:2020dww}:
\begin{eqnarray}
	\Delta a_e = a_e^{\text{exp}} - a_e^{\text{SM}} = (4.8\pm 3.0)\times 10^{-13}.\label{eqn:electrongm2}
\end{eqnarray}
On the other hand, the recently published result of the Fermilab-based Muon $g-2$ experiment~\cite{Muong-2:2021ojo},
when combined with that of the E821 collaboration at
BNL~\cite{Muong-2:2006rrc}, has an excess of about $4.2\sigma$ over the latest SM theoretical computation~\cite{Aoyama:2020ynm}:
\begin{eqnarray}
\Delta a_\mu \equiv  a_\mu^{\text{exp}} - a_\mu^{\text{SM}} = (25.1\pm 5.9)\times 10^{-10}.\label{eqn:muongm2}
\end{eqnarray}
The discrepancy may be explained by an ALP~\cite{Cornella:2019uxs,Buen-Abad:2021fwq,Keung:2021rps}.
In principle, both diagonal and off-diagonal couplings could lead to new contributions to $a_\alpha=(g-2)_\alpha/2$ ($\alpha=e,\mu$).
However, since in the present work we keep $g_{\mu e}$ switched off and also take equal scalar and
pseudoscalar couplings, the LFV couplings contributions vanish~\cite{Cornella:2019uxs}.
The diagonal coupling  contributions to $a_l$ are
\begin{eqnarray}
(\Delta a_{l_\alpha})_{\text{diag.}}= -\frac{m_{\alpha}^2}{4\pi^2} \Big[  16\pi\, \alpha_{\text{em}}\, g_{\gamma\gamma}^{\text{eff}}\, g_{\alpha\alpha}\Big(  \text{log}\frac{\Lambda^2}{m_\alpha^2} - h_2(x_\alpha)  \Big)   +|g_{\alpha\alpha}|^2 h_1(x_\alpha)   \Big]  ,\label{eq:deltaa}
\end{eqnarray}
where $x_\alpha=m_X^2/m^2_{\alpha}$, $g_{\gamma\gamma}^{\text{eff}}$ can be computed by Eq.~\eqref{eq:decaywidthX2gg}, and
\begin{eqnarray}
h_1(x)&=&1+2x-(x-1)x\,\text{log}x+2x(x-3)\sqrt{\frac{x}{x-4}}\,\text{log}\Big(\frac{\sqrt{x}+\sqrt{x-4}}{2}\Big),\\
h_2(x)&=&1+\frac{x^2}{6}\text{log}x-\frac{x}{3}-\frac{x+2}{3}\sqrt{(x-4)x}\,\text{log}\Big( \frac{\sqrt{x}+\sqrt{x-4}}{2} \Big).
\end{eqnarray}
In Eq.~\eqref{eq:deltaa},  the second term, $-\frac{m^2_{\alpha}}{4\pi^2}|g_{\alpha\alpha}|^2h_1(x_\alpha)$, is strictly negative.
Therefore, in order to explain the observed discrepancy, the first term in Eq.~\eqref{eq:deltaa} must be positive and large enough  to overcome the negative second term.
We find that for the electron $g-2$ case, the leptophilic axions considered in Scenario 1 can bring the theoretical prediction to reach the 95\% lower limit of $\Delta a_e$ but not up to the central value.
On the other hand, for the muon $g-2$ discrepancy the ALPs in Scenario 2 cannot even bring the theoretical prediction to the edge of the 95\% lower limit of the experimental value (cf.~Eq.~\eqref{eqn:muongm2}).
For these reasons, we do not show their corresponding limits in the sensitivity plots in Sec.~\ref{sec:results}.
Lastly, we note that a very recent lattice calculation of hadronc contributions to the muon $g-2$ pushed the
   SM prediction to within $1\sigma$ of the experimental result \cite{Borsanyi:2020mff}.

\subsection{Beam-dump experiments and dark-photon searches}

\begin{figure}[htb]
	\centering
	\includegraphics[width=0.49\textwidth]{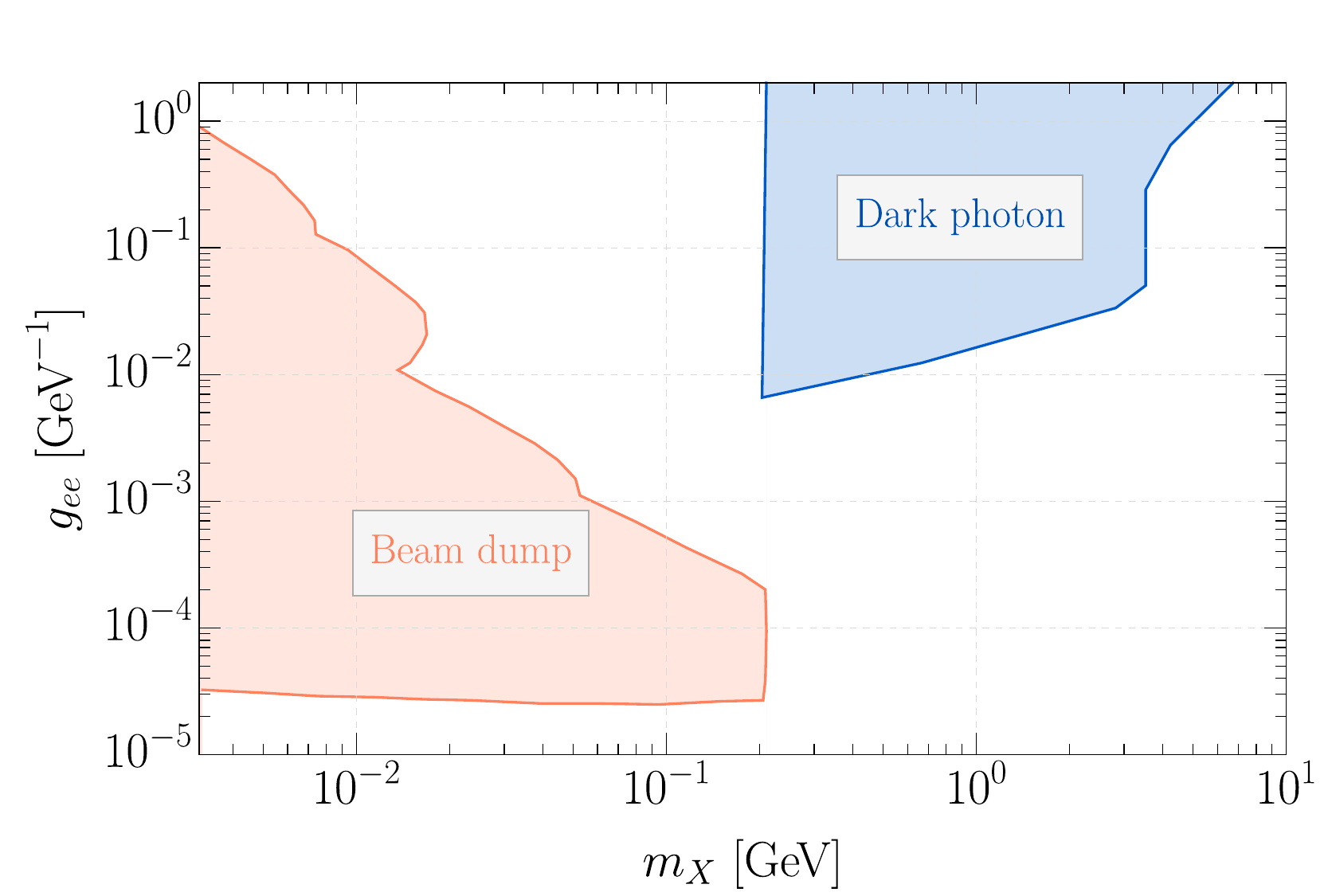}
	\caption{ Constraints on $g_{e e}$ as a function of $m_X$,
          reproduced from Ref.~\cite{Bauer:2017ris}. Note that the BaBar dark photon bound is valid for universal diagonal couplings $g_{ee}=g_{\mu\mu}=g_{\tau\tau}$.
	}
	\label{fig:BD_BaBar}
\end{figure}

In our $m_X$ range of interest, results from the E137 and E141 beam-dump experiments at SLAC~\cite{Riordan:1987aw}, and similar ones at CHARM~\cite{CHARM:1985anb}, Orsay~\cite{Davier:1989wz}, and Fermilab~\cite{Bross:1989mp}, typically constrain the coupling $g_{ee}$ to be outside the range $10^{-5}$ GeV$^{-1}$ to 1 GeV$^{-1}$ when taking the axion to radiate off the electron beam~\cite{Essig:2010gu,Andreas:2010ms}.
Moreover, assuming universal diagonal couplings $g_{ee}=g_{\mu\mu}=g_{\tau\tau}$, a dark-photon search  performed at BaBar~\cite{BaBar:2016sci,Bauer:2017ris} can be recast into bounds on these couplings~\cite{Bauer:2017ris}.
The limits obtained exclude $g_{ee}$ above $\sim 6\times 10^{-3}$ GeV$^{-1}$ for $m_X$ between about 210 MeV and 8 GeV.\footnote{The parameter $c_{ee}^{\text{eff}}/\Lambda$ used in Ref.~\cite{Bauer:2017ris} is related to $g_{ee}$ of the present work by a factor 2: $c_{ee}^{\text{eff}}/\Lambda=2g_{ee}$.}
We extract these results from Ref.~\cite{Bauer:2017ris} and show them in Fig.~\ref{fig:BD_BaBar}.

\subsection{Muonium-antimuonium oscillations and $\mu^-\to e^-$ conversion}

The bound state of $e^-$ and $\mu^+$ is known as muonium $M$.
In the presence of lepton flavor violation, the muoniums may oscillate into antimuoniums $\bar{M}$ ($e^+ \mu^-$).
The strongest bound on the probability of spontaneous muonium to antimuonium conversion, $P_{M \bar{M}}<8.3 \times 10^{-11}$ at 90\% C.L.,  was obtained by the MACS experiment in 1999~\cite{Willmann:1998gd}.
For the leptophilic ALPs, only the coupling $g_{\mu e}$ could result in such an oscillation.
Since we keep this coupling off in this work, we do not go into the details of the derivation of the bound.
	
Similarly, $\mu^-\to e^-$ conversion in nuclei is also a very sensitive probe of LFV~\cite{Calibbi:2017uvl}.
In the context of the leptophilic ALPs, this process also only concerns the coupling $g_{\mu e}$, and is not studied here.
Bounds derived from past experiments~\cite{SINDRUMII:2006dvw} and those projected for
future experiments~\cite{Mu2e:2014fns,COMET:2018auw} can be found in Ref.~\cite{Cornella:2019uxs}.

%% file: tex/simulation.tex

\section{Prompt and displaced searches at Belle~II}\label{sec:simulation}
In this section we discuss our recast of the Belle prompt LFV search, projected to the Belle~II sensitivity, as well as the displaced search that we propose to conduct.
Our results are then shown in Section~\ref{sec:results}.

Belle II~is an ongoing $B$-factory experiment hosted at the KEK laboratory in Tsukuba, Japan.
It collides an electron beam of energy $E_{e^-}=7$ GeV with a positron beam of $E_{e^+}=4$ GeV, reaching a center-of-mass energy of 10.58 GeV.
The experiment has so far collected about 200~fb$^{-1}$, less than the samples collected by the BaBar and Belle experiments, approximately 500~fb$^{-1}$ and 1~ab$^{-1}$, respectively.
However, Belle~II is planned to eventually collect a sample of 50~ab$^{-1}$. 
With a cross section $\sigma(e^+ e^-\to\tau^+ \tau^-)\approx 0.92$~nb, this implies $4.6\times 10^{10}$ $\tau$-pair events produced.

\subsection{Recast of the prompt search}

We first recast the Belle prompt search for lepton-flavor-violating $\tau$ decays into three charged leptons, which used a sample of $7.19\times 10^8$ $\tau$-pair events~\cite{Hayasaka:2010np}.

The background level reported in Ref.~\cite{Hayasaka:2010np} ranges from $0.01\pm 0.01$ events in the $\tau^-\to \mu^+e^-e^-$ channel to $0.20\pm 0.15$ events in the $\tau^-\to e^+e^-e^-$ channel.
We assume that the number of background events will remain small even with the much larger sample of Belle~II, sufficiently so that it can be ignored in our Belle~II projection.

\begin{table}[htb]
	\centering
	\begin{tabular}{c|c}
	Decay modes	& Baseline efficiency  	\\
		\hline
		$\tau\to X e, X\to e^- e^+$	 		& $6.0\,\%$ \\
		$\tau\to X \mu, X\to \mu^- \mu^+$	& $7.6\,\%$ \\
	\end{tabular}
	\caption{Baseline efficiencies reproduced from Ref.~\cite{Hayasaka:2010np}.}
	\label{tab:baseline_efficiency}
\end{table}
The event-selection efficiency reported in Ref.~\cite{Hayasaka:2010np} for each final state is reproduced in Table~\ref{tab:baseline_efficiency}.
We refer to this as the ``baseline efficiency". 
In addition, since the ALP in the region of sensitivity can be long-lived, its daughter leptons may escape detection in the analysis optimized for prompt decays, reducing the efficiency.
We use events simulated with Pythia8.245~\cite{Sjostrand:2014zea} and without detector simulation to estimate the efficiencies as functions of the ALP lifetime and mass.
While the beams at Belle~II collide with a small crossing angle, we simplify this and simulate head-on collisions, with the electron beam momentum in the $+z$ direction.

As required in Ref.~\cite{Hayasaka:2010np}, we require  
the transverse and longitudinal impact parameters of both charged leptons produced in the ALP decay to satisfy $d_0<5\text{ mm}$ and $z_0<30\text{ mm}$, respectively.
The computation of $z_0$ is performed with
\begin{eqnarray}
z_0 = \left|z - \frac{(x\cdot p_x+y\cdot p_y)\cdot p_z}{p_T^2}\right|,\label{eq:z0}
\end{eqnarray}
where $x,y,$ and $z$ are the production position coordinates of the charged lepton with respect to the IP, $p_x, p_y,$ and $p_z$ are the components of the lepton momentum in the three spatial directions, and $p_T^2=p_x^2+p_y^2$ is its transverse momentum squared.

The calculation of the transverse impact parameter is more involved.
For straight tracks, it can be approximated as
\begin{eqnarray}
d_0^{\text{naive}} = \frac{|p_x\cdot y -p_y \cdot x|}{p_T}.
\end{eqnarray}
However, given the relatively low momenta and possibly large flight distances, the track curvature resulting from the local magnetic field can be non-negligible.
We estimate the radius of the track $R$ as
\begin{eqnarray}
R\simeq \frac{1}{0.3}\frac{p_T}{\text{GeV}} \frac{\text{T}}{B} \text{ meters},
\end{eqnarray}
where $\text{T}$ stands for Tesla and $B=1.5\text{ T}$ is the strength of the magnetic field.
The true transverse impact parameter $d_0$ is then computed with 
\begin{eqnarray}
d_0 = \sqrt{(R+d_0^{\text{naive}})^2+r^2-(d_0^{\text{naive}})^2}-R,
\end{eqnarray}
where $r$ labels the transverse decay position of the ALP from the interaction point (IP).

Furthermore, we require $r<10\text{ cm}$.
This requirement does not appear explicitly in Ref.~\cite{Hayasaka:2010np}, but we apply it in order to obtain a more realistic simulation of the minimal number of detector hits needed for track reconstruction.
This requirement is important only for very light ALPs, which are highly boosted and can pass the $d_0$ and $z_0$ requirements even when decaying far from the IP.
It is redundant for heavy ALPs, whose daughters are separated by a large angle and would anyway fail the $d_0$ and $z_0$ cuts.

\subsection{Proposal of a displaced-vertex search}

Since the prompt search is not optimized for detection of a long-lived ALP, we propose to complement it with a displaced search, in which the maximal $d_0$ and $z_0$ requirements of the prompt analysis are replaced with a minimal requirement on the ALP flight distance, $r>1$~cm.
Since this displaced-vertex requirement leads to a large reduction in background (see, e.g., Ref.~\cite{Lee:2018pag}), the background in this search is even smaller than the negligible background in the prompt search.

Given the strong background suppression provided by the displaced vertex and the sub-event background level of the prompt Belle analysis~\cite{Hayasaka:2010np}, we expect that many of the cuts applied in the prompt analysis can be avoided in the displaced one.
For example, based on the efficiencies reported in Ref.~\cite{Hayasaka:2010np}, removing the missing-momentum cut and applying lepton identification criteria on only one of the leptons increases the baseline efficiency by a factor of 1.5 in scenario~1 and 2.3 in scenario~2.
These efficiency enhancements relative to the prompt analysis are included into the final sensitivity estimates for the displaced search reported in Section~\ref{sec:results}.

In addition to the baseline efficiency, we determine the detection efficiency variation as a function of the ALP decay position following the procedure used in Refs.~\cite{Dib:2019tuj, Dey:2020juy}.
We require the ALP to decay inside the fiducial volume of the tracker, defined by the transverse and longitudinal distances from the IP: $1\text{ cm} < r < 80\text{ cm}$ and $-40\text{ cm} <z < 120\text{ cm}$, respectively.
In addition, we account for the tracking efficiency decrease with the transverse distance from the IP.
This so-called ``displaced-tracking efficiency" $\epsilon^{\text{track}}$ is parameterized as a linear function of $r$, so that $\epsilon^{\text{track}}=1$ for $r=1$~cm and $\epsilon^{\text{track}}=0$ at $r=80$~cm.
In particular, in the large decay length limit, $\epsilon^{\text{track}}$ is effectively reduced to an overall factor of $50\%$ on average, for decays inside the fiducial volume.

\subsection{Computation and Simulation}

The number of signal events at Belle~II is computed with:
\begin{eqnarray}
N_S^{\text{Belle II}}&=&2\cdot N_{\tau^-\tau^+} \cdot \text{BR}(\tau\to\text{1 prong})    \nonumber\\
&& \cdot \text{BR}(\tau \to X l_\alpha)\cdot \epsilon\cdot \text{BR}(X\to l_\alpha^-  l_\alpha^+),\label{eq:NS}
\end{eqnarray}
where $l_\alpha=e$ or $\mu$, $N_{\tau^-\tau^+}=4.6\times 10^{10}$ is the total number of $\tau$ pair events produced at Belle~II,  $\epsilon$ denotes the final efficiency of event selections as detailed above, and BR$(\tau\to\text{1 prong})\approx 85\%$ is the SM decay branching ratio of the $\tau$ lepton into a single charged particle and neutrinos. Application of this factor assumes that the events are identified as $e^+e^-\to \tau^+\tau^-$ by requiring the topology of one track recoiling against the three leptons from the signal decay chain.
The factor 2 is due to the fact that in each signal event there are two $\tau$ leptons.
For vanishing background, $N_S^{\text{Belle II}}=3$ corresponds to the $95\%$ C.L. exclusion limits.

The simulations are performed on a grid in the model parameters.
For each grid point, we simulate the process $e^-e^+\to \tau^-\tau^+$.
The $\tau$ leptons decay to $X+e$ or $X+\mu$, and the ALP decays to $e^- e^+$ or $\mu^- \mu^+$, respectively.
Pythia automatically processes the decay positions of the simulated ALPs according to the exponential decay distributions using the ALP lifetime and boost factor.
Then the prompt and displaced sets of event selection criteria are imposed at the parton level to obtain the efficiency 
separately for each of the two search strategies.
Thus, we obtain the efficiencies for the two strategies at each grid point.

%% file: tex/results.tex

\section{Numerical Results}\label{sec:results}

We now proceed to present numerical results in this section.
We concentrate on the two representative benchmark scenarios introduced in Sec.~\ref{sec:model} for detailed numerical studies in Secs.~\ref{subsec:scenario1} and~\ref{subsec:scenario2} , while for other further possibilities we will provide a qualitative discussion in Sec.~\ref{subsec:other_scenarios}.
Since, as explained in Sec.~\ref{sec:simulation}, for both prompt and displaced searches we expect zero background events, we use 3-signal-event isocurves to show exclusion
limits at $95\%$ C.L.

\subsection{Scenario 1: $g_{\tau e}$ and $g_{ee}$}\label{subsec:scenario1}

In Scenario 1, the couplings $g_{\tau e}$ and $g_{ee}$ are turned on, mediating the production and decay of the ALP, respectively, while the other couplings are all considered as vanishing. Thus, the signal events are $\tau\to X e$ with $X\to e^-  e^+$.

\begin{figure}[t]
	\centering
	\includegraphics[width=0.49\textwidth]{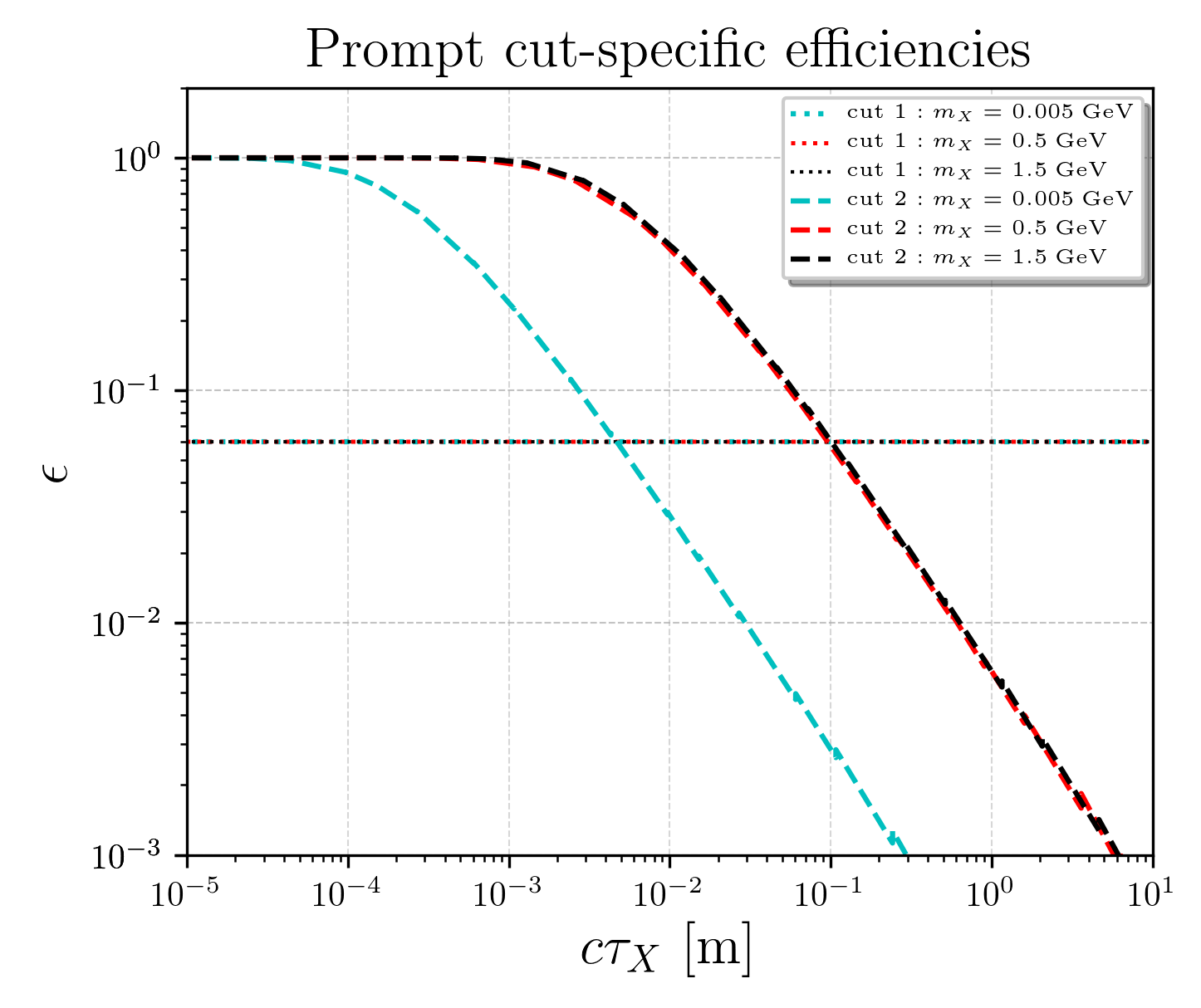}	\includegraphics[width=0.49\textwidth]{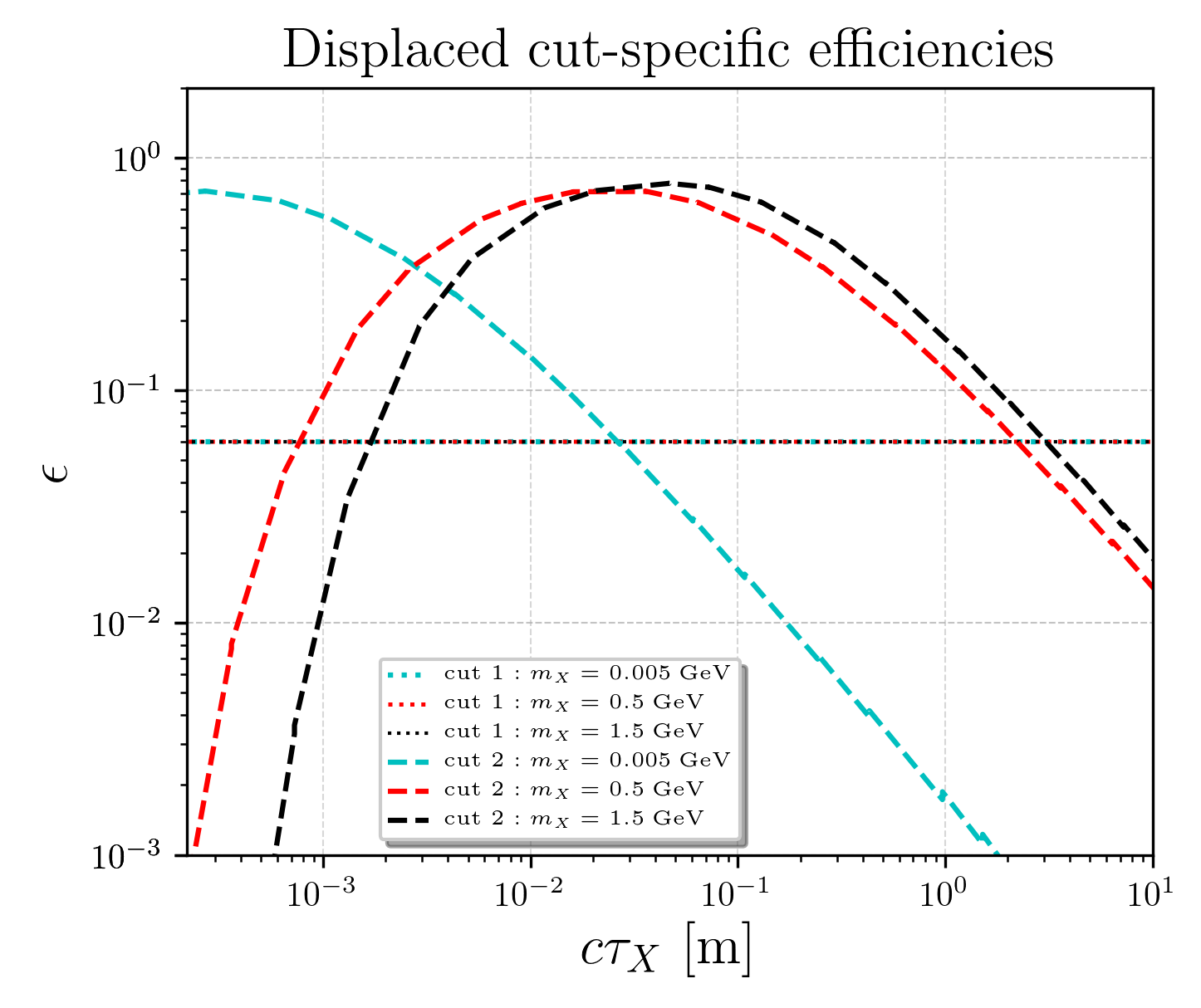}	\caption{Efficiencies vs.~$c\tau_X$ for the signal events of Scenario 1, with three benchmark values of $m_X$: 0.005, 0.5, 1.5 GeV.
		\textit{Left}: 	 prompt cut-specific efficiencies.
		\textit{Right}: displaced cut-specific efficiencies.
	``Cut 1" denotes the baseline efficiency. For the prompt search, ``cut 2" is the efficiency of requiring $d_0<5\text{ mm}$, $z_0<30\text{ mm}$, and $r<10\text{ cm}$.
	For the displaced search, ``cut 2" gives the efficiency of applying the fiducial-volume and the displaced-tracking requirements.
			}
	\label{fig:efficiencies_vs_ctau_scenario1}
\end{figure}
We show in Fig.~\ref{fig:efficiencies_vs_ctau_scenario1} the cut-specific event selection efficiencies for the signal events as functions of $c\tau_X$ for the prompt and displaced analysis and for three benchmark masses $m_X=0.005$ (blue), 0.5 (red), and 1.5 (black) GeV.

\begin{figure}[t]
	\centering
	\includegraphics[width=0.49\textwidth]{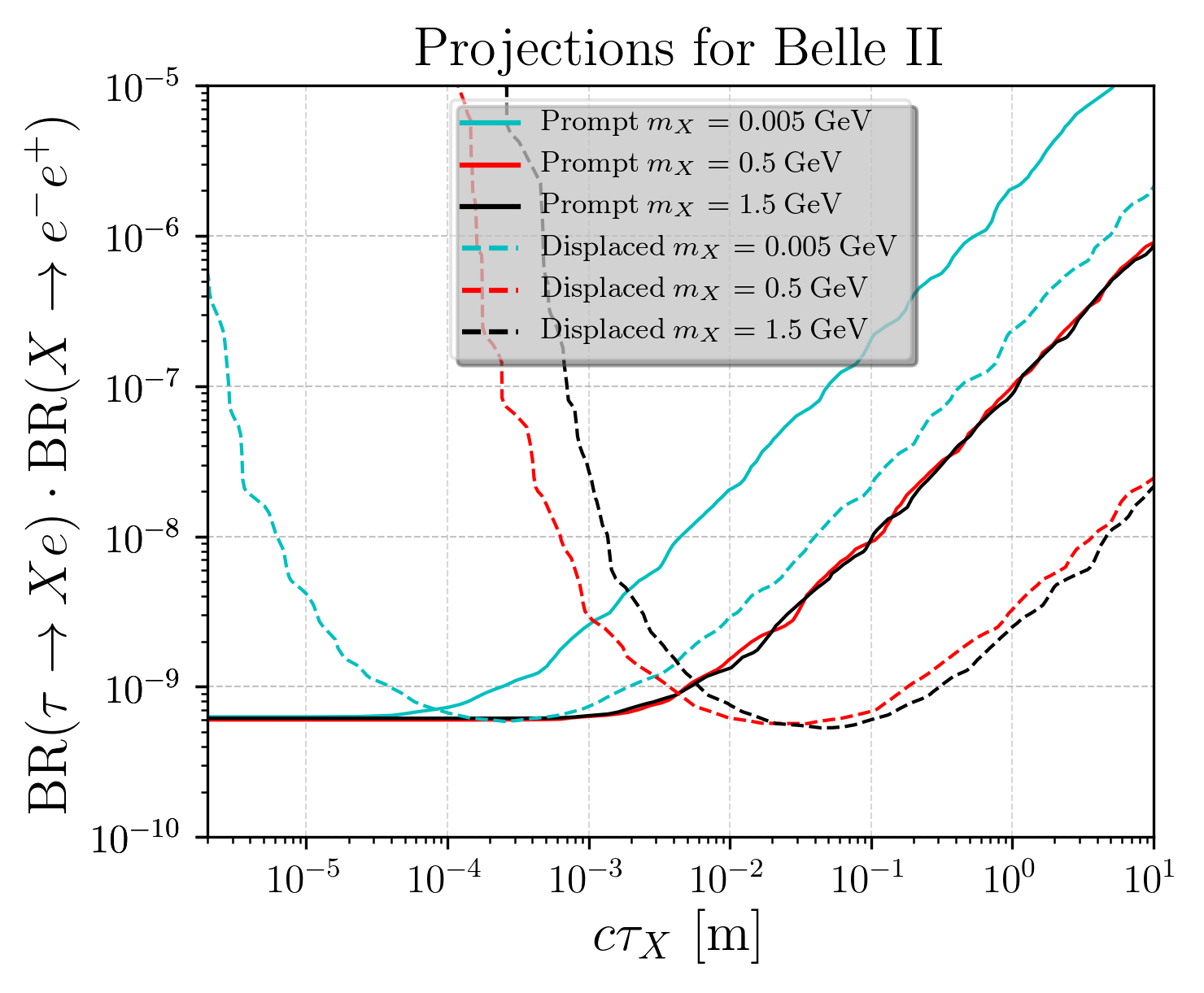}\\
	\includegraphics[width=0.49\textwidth]{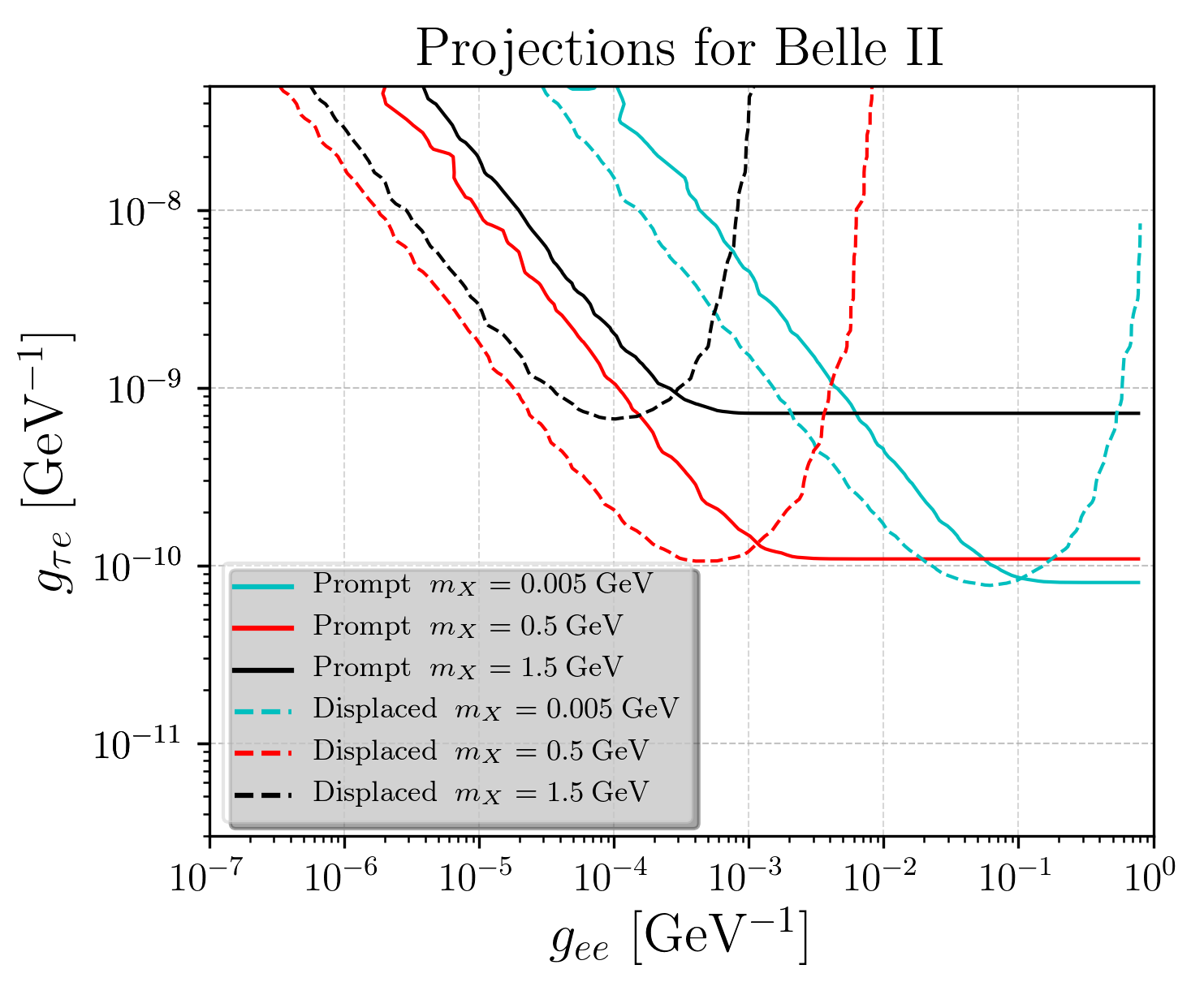}
	\includegraphics[width=0.49\textwidth]{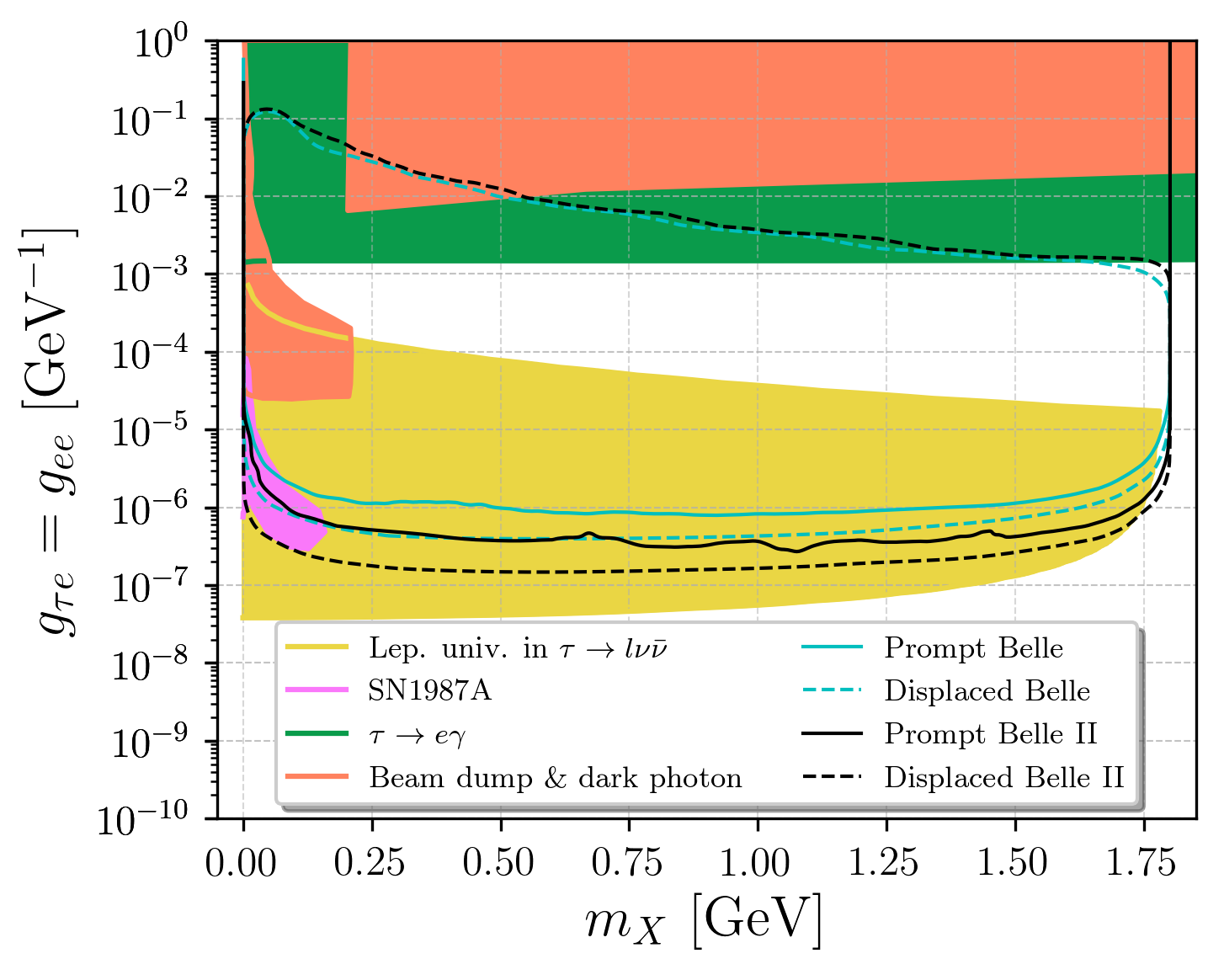}
	\caption{Sensitivity results for Scenario 1 at 95\% C.L.
		\textit{Top}: model-independent limits on the branching-fraction product BR$(\tau\to X e)\cdot$BR$(X\to e^- e^+)$ vs.~$c\tau_X$ for various ALP masses.
		\textit{Bottom left}: model-dependent bounds from Belle and Belle~II on $g_{\tau e}$ vs. $g_{ee}$ for the prompt and displaced searches at Belle~II for various ALP masses.
		\textit{Bottom right}: bounds on the couplings vs. ALP mass in the specific case $g_{\tau e}=g_{ee}$, compared with bounds from lepton universality in $\tau\to l\nu \bar{\nu}$~\cite{Pich:2013lsa,BaBar:2009lyd}, $\tau\to e\gamma$~\cite{BaBar:2009hkt}, SN1987A~\cite{Lucente:2021hbp}, beam-dump experiments~\cite{Essig:2010gu,Andreas:2010ms}, and a dark-photon search~\cite{BaBar:2016sci,Bauer:2017ris}.
		Note that the lepton-universality constraint is only valid for long-lived ALPs, resulting in the upper cutoff of the corresponding region.
	}
	\label{fig:sensitivities_scenario1}
\end{figure}
Fig.~\ref{fig:sensitivities_scenario1} presents the estimated 95\% C.L. sensitivities at Belle~II for the prompt and displaced searches.
The upper plot shows the projected limits on the branching fraction product BR$(\tau\to X e)\cdot$BR$(X\to e^+ e^-)$ as a function of $c\tau_X$ for the three benchmark ALP masses.
Since both searches have no background, their maximal sensitivities are approximately equal, with the prompt (displaced) search being more sensitive at shorter (longer) ALP lifetimes.
Focusing our attention on the displaced search, which is proposed here for the first time, we observe that it is up to about 40 times more sensitive to the branching-fraction product than the prompt search at large values of $c\tau_X$.
This translates to about a factor of 6 in the sensitivity to $g_{\tau e}$ for a given value of $g_{ee}$, as seen in the lower-left plot. 
We note that the next-tightest constraint, which comes from lepton universality in $\tau\to l \nu \bar{\nu}$ (see Section~\ref{subsec:lepton_universality}), yields $g_{\tau e}\lesssim 10^{-7}$~GeV$^{-1}$ (which is valid only for long-lived ALPs and hence small $g_{ee}$ coupling).
Since this is much weaker than the sensitivity reach of the prompt and displaced searches, it is not shown on the same plot.

In the lower-right plot we study the particular case $g_{\tau e}=g_{ee}$, and present the limits on the coupling value vs.~$m_X$ for both Belle~II and Belle luminosities, together with bounds from existing measurements.
Since the $\tau\to l\nu\bar\nu$ lepton-universality bound is only valid for long-lived ALPs, we show the corresponding excluded parameter space only for $c\tau_X\gtrsim 1$ m.
 In this case, the tightest bound in the case of small couplings is obtained from the lepton-universality constraint. 
 However, the prompt $\tau\to 3\ell$ and the displaced search proposed here are uniquely sensitive to a mass-dependent range of coupling values between about $2\times 10^{-5}$ and $2\times 10^{-3}$.

We note that our ``model-independent" limits on the branching-fraction product can be used to obtain constraints on other models with similar scattering and decay topologies. Examples include a light new gauge boson $Z'$~\cite{Heeck:2016xkh,Chen:2017cic,Foldenauer:2016rpi} or a light $CP$-even scalar~\cite{Flacke:2016szy,BhupalDev:2016nfr,Dev:2017dui}.

\subsection{Scenario 2: $g_{\tau \mu}$ and $g_{\mu\mu}$}\label{subsec:scenario2}

In Scenario 2, the non-zero couplings $g_{\tau \mu}$ as well as $g_{\mu\mu}$ are responsible for the production and decay of the ALPs, respectively.
The sensitive mass range is now between $2\, m_\mu$ and $m_\tau-m_\mu$.

\begin{figure}[t]
	\centering
	\includegraphics[width=0.49\textwidth]{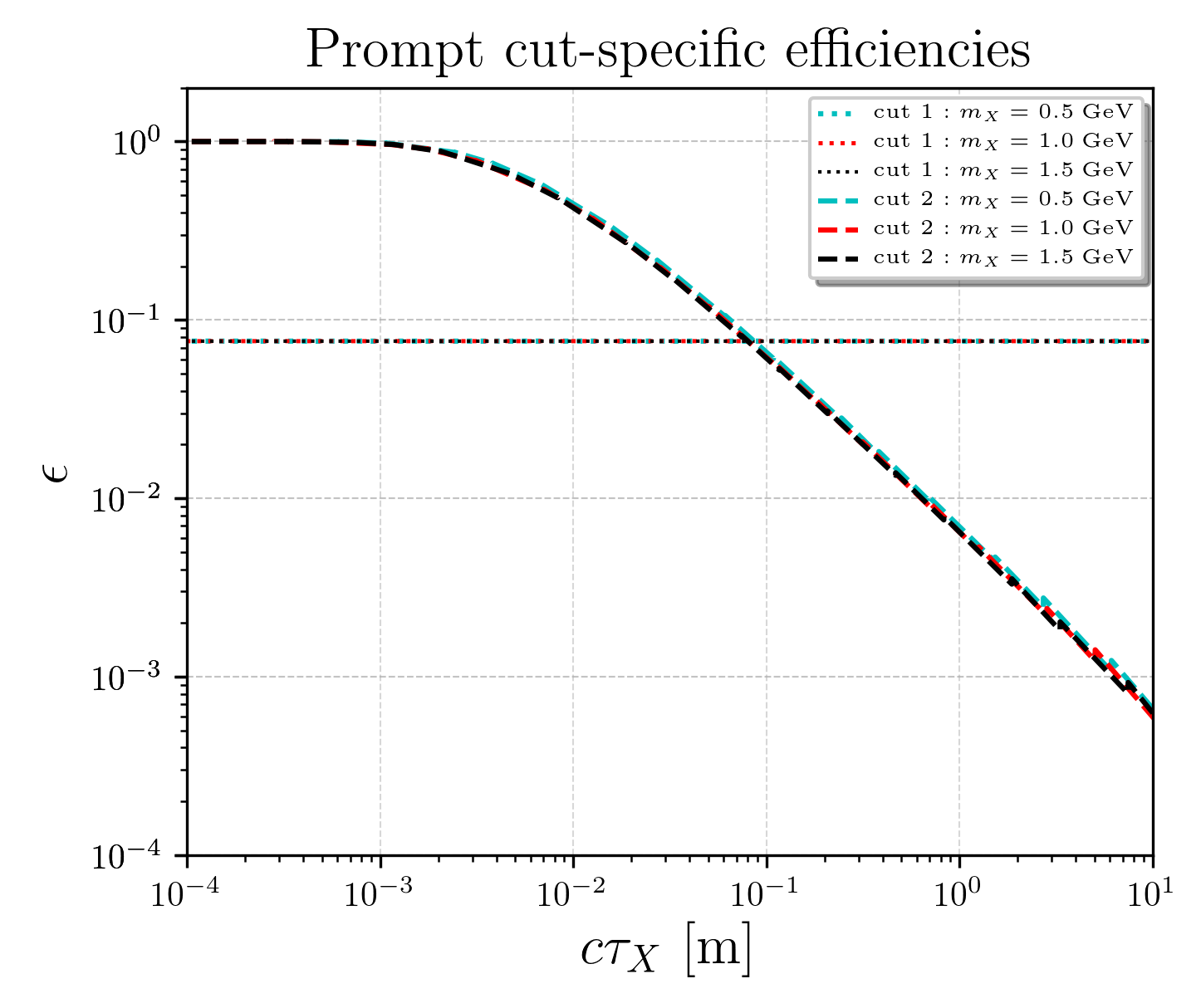}	
	\includegraphics[width=0.49\textwidth]{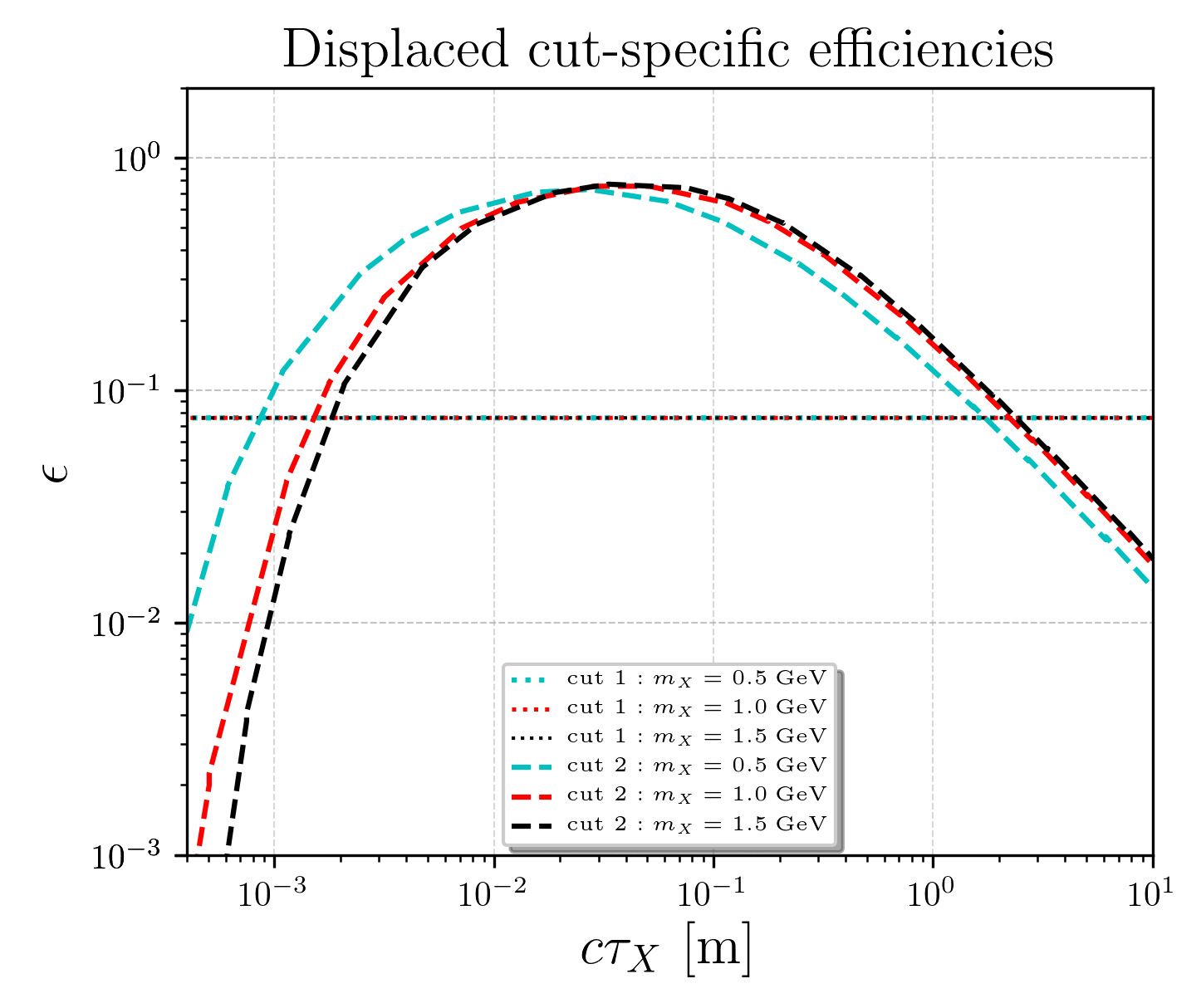}
	\caption{Efficiencies for Scenario 2 ($g_{\tau \mu}$ and $g_{\mu \mu}$ are non-zero). The format is the same as in Fig.~\ref{fig:efficiencies_vs_ctau_scenario1}, except for the different set of mass choices: $m_X=0.5$, 1.0, and 1.5 GeV.
	}
	\label{fig:efficiencies_vs_ctau_scenario2}
\end{figure}
In Fig.~\ref{fig:efficiencies_vs_ctau_scenario2} we show the signal event selection efficiencies as functions of $c\tau_X$, for $m_X=0.5$, 1.0, and 1.5 GeV.
These plots are similar to those of Fig.~\ref{fig:efficiencies_vs_ctau_scenario1}, except for some differences that mainly stem from the different mass choices, as well as different baseline efficiencies.

\begin{figure}[t]
	\centering
	\includegraphics[width=0.49\textwidth]{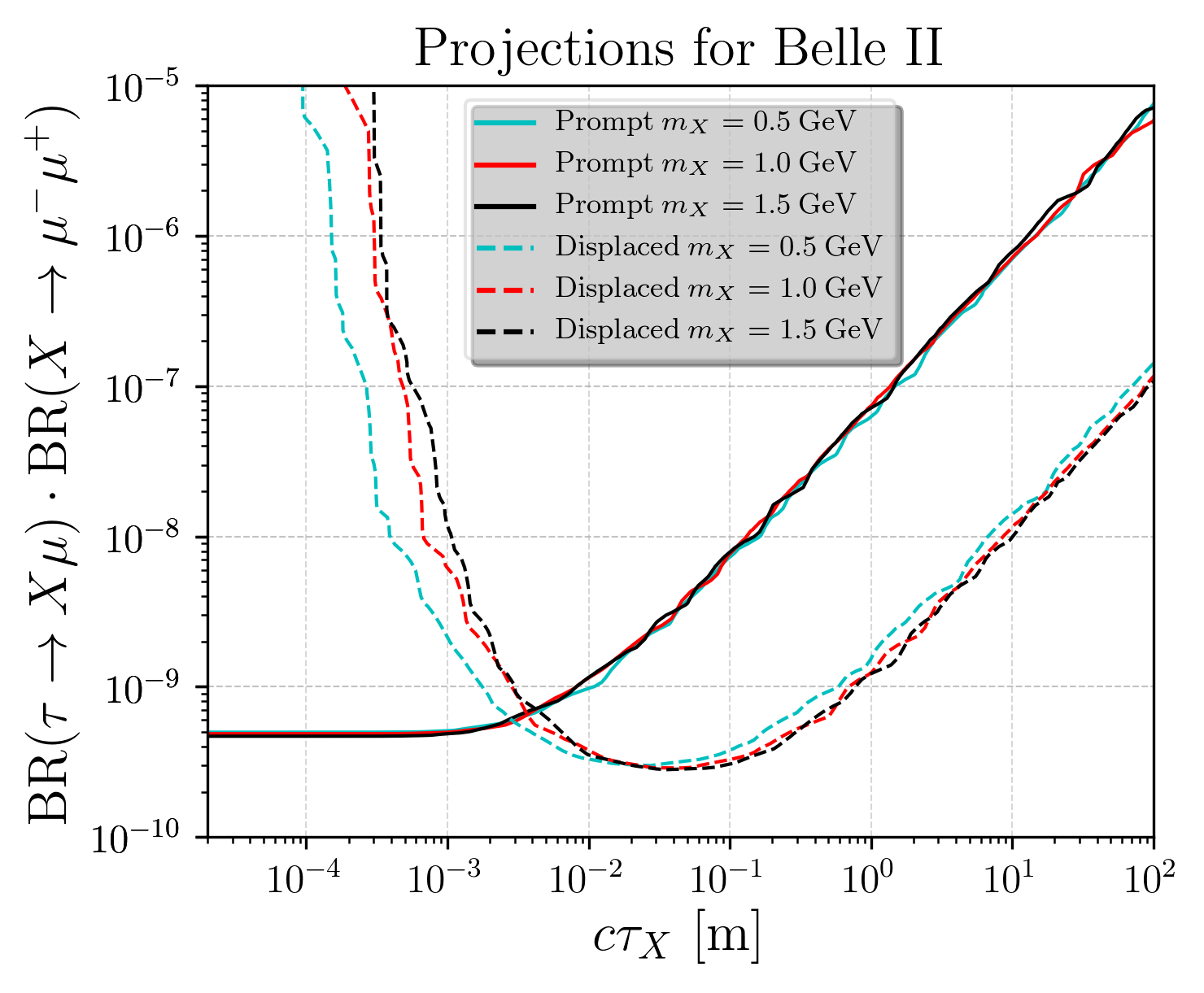}\\
	\includegraphics[width=0.49\textwidth]{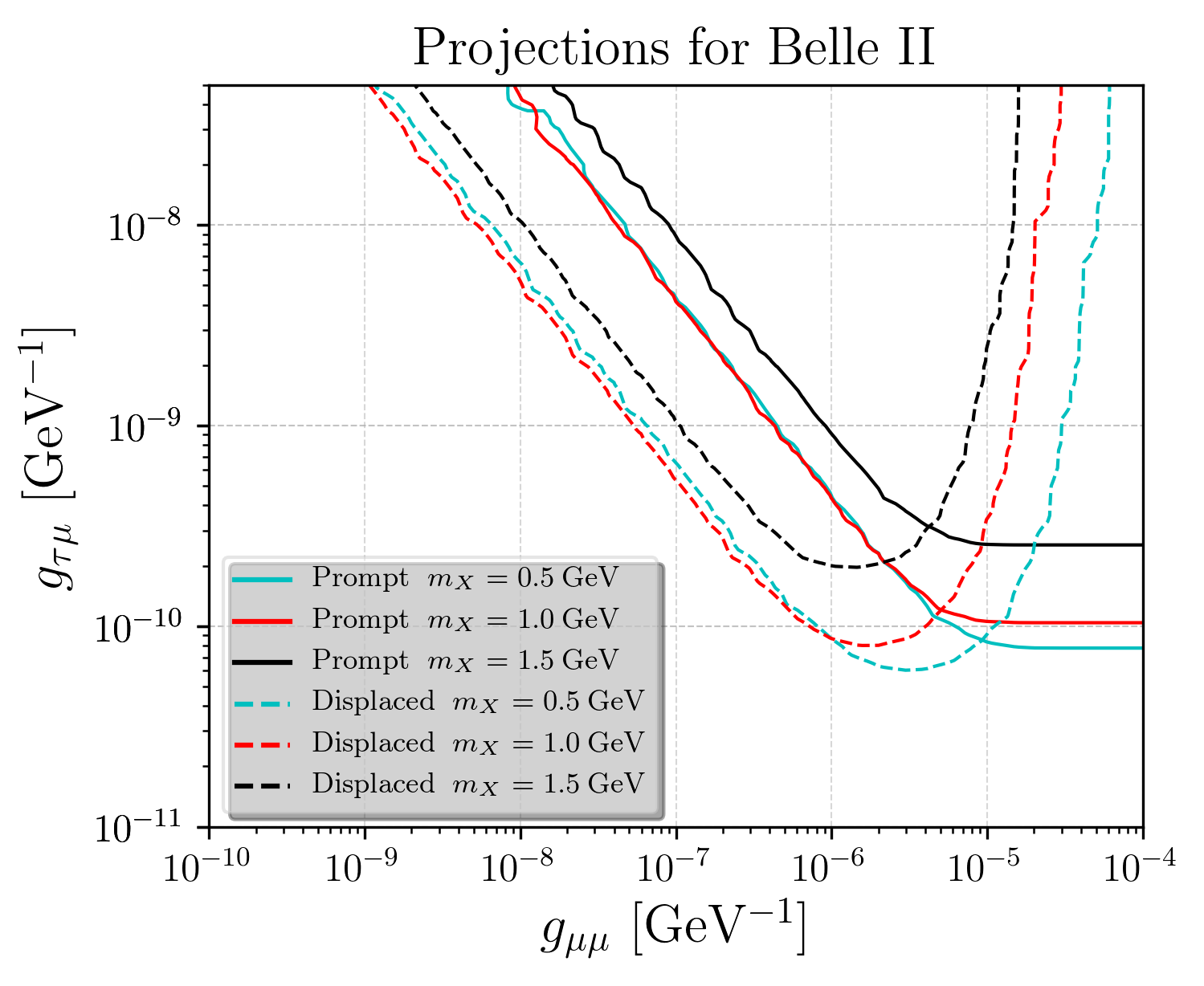}
	\includegraphics[width=0.49\textwidth]{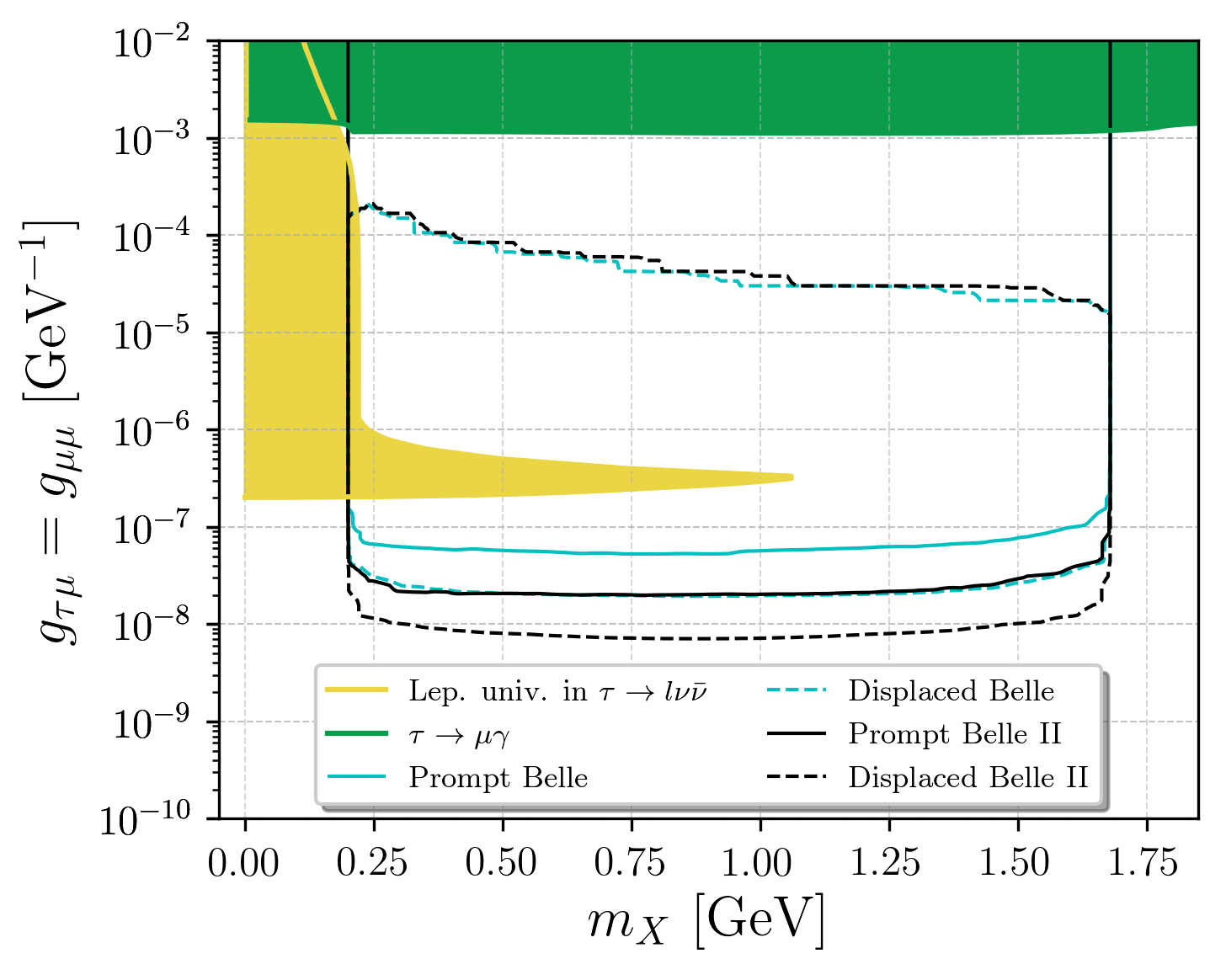}
	\caption{Sensitivity results for Scenario 2 ($g_{\tau \mu}$ and $g_{\mu \mu}$ are non-zero). The format is the same as in Fig.~\ref{fig:sensitivities_scenario1}, except for the different mass choices.
	}
	\label{fig:sensitivities_scenario2}
\end{figure}
We present the final sensitivity estimates for Scenario~2  in Fig.~\ref{fig:sensitivities_scenario2} with the same format as in Fig.~\ref{fig:sensitivities_scenario1}.
Due to the similar efficiencies of the two scenarios, the branching-fraction bounds are similar.
However, the $g_{\mu\mu}$ sensitivity of scenario~2 is significantly enhanced, mainly because the width $\Gamma(X\to ll)$ is proportional to the lepton mass squared.
As a result, in the case $g=g_{\tau\mu}=g_{\mu\mu}$ shown in the bottom-right plot, the displaced search can probe $g$ values that are smaller by more than a factor of 10 than those of the lepton-universality constraint.
The measurement of $\Gamma( \tau \to \mu \nu \bar{\nu})/\Gamma( \tau \to e \nu \nu) $ is larger than	the SM expectation.
Therefore, the presence of $g_{\tau \mu}$ and $g_{\mu\mu}$  couplings will bring the prediction closer to the central value of the measurement, so that the valid parameter space satisfies $R_{\mu e}^{SM +X} / R_{\mu e}^{SM}  - 1 < 0.0072 + 2 \times 0.004$.
In contrast to the case of Scenario 2, the presence off $g_{\tau e}, g_{ee}$ in Scenario 1 will bring the prediction further away from the central value of the measurement, thus resulting in a more stringent constraint.
The valid parameter space has to satisfy $R_{e \mu }^{SM +X} / R_{e \mu}^{\text{SM}}- 1 <  - 0.0072 + 2 \times 0.004$. 
It implies that the region excluded in Scenario 1 is much larger than that of Scenario 2.

We note that, while we focused on $X\to \mu^+\mu^-$ decays that take place inside the tracker fiducial volume, the muon tracks can also be found when the decay takes place in the muon detector, which extends out to a distance of about 3.5~m from the IP. Naively, the somewhat larger decay distance allows for a somewhat stronger reach. However, since the muon detector has no magnetic field, it is impossible to determine the momenta of the muons in this case. This would likely result in high background levels that would make this approach less sensitive.

\subsection{Other Scenarios}\label{subsec:other_scenarios}

We discuss briefly some other coupling scenarios. 

One possibility is to take the combinations $(g_{\tau e},g_{\mu \mu})$ or $(g_{\tau \mu},g_{e e})$ to be non-zero.
However, these results would be rather similar to those of Scenario 2 and Scenario 1, respectively, except at the edges of phase space.
This is because the production of the ALP is dominantly proportional to $m_\tau^3$, and is not affected by the light lepton or $X$ masses except at the kinematic threshold (\textit{cf}.~Eq.~\eqref{eq:taudecaywidth}).
Another possibility is to take both $g_{\tau e}=g_{\tau \mu}$ and $g_{ee}=g_{\mu\mu}$ to be non-vanishing.
In this case the ALP decay is dominated by the di-muon channel, because $m_{\mu}\gg m_e$, and the production rate is approximately doubled compared to that in Scenarios 1 and 2 with the same LFV coupling values.
Therefore, the end results are similar to those of Scenario 2 presented here, except for a factor of~2 in the production rate.
It is nevertheless worth mentioning that LFV muon decays such as $\mu\to e\gamma$ can place strong constraints on this scenario.

We also consider the possibility of a non-zero $g_{e\mu}$. According to Eqs.~\eqref{eq:decaywidthX2ll2} and~\eqref{eq:decaywidthX2ll}, equal values of $g_{e\mu}$ and $g_{\mu\mu}$ lead to equal widths for $X\to e^\pm \mu^\mp$ and $X\to \mu^+\mu^-$ except near the kinematic thresholds.
Therefore, we expect the sensitivity for $g_{e\mu}$, together with, e.g., non-vanishing $g_{\tau \mu}$, to be similar to that obtained for Scenario 2.
We note that LFV muon-decay processes such as $\mu\to e+$ missing can strongly constrain $g_{e\mu}$ for $m_X < m_\mu$.

Finally, if the coupling $g_{\tau\tau}$ is non-zero, it would lead to ALP-strahlung production of the ALP by radiation off the $\tau$ (see Ref.~\cite{BaBar:2020jma} for an experimental search at BaBar with this production mechanism)  and would loop-induce the decay $X\to \gamma\gamma$ for the $m_X$ range considered in this work.

%% file: tex/conclusions.tex

\section{Conclusions}\label{sec:conclusions}

In this paper we investigated the sensitivities of the $B$-factory experiment Belle~II to a leptophilic axion-like particle $X$ with lepton-flavor-violating (LFV) couplings.
The decay channels of interest are $\tau \to X  l$, where $l=e,\mu$, with the ALP undergoing the decay $X\to l^+ l^-$. 
The ALP can also decay to a pair of photons via a lepton loop.
We do not consider this decay as a search signature, but we do account for it when calculating the signal branching fractions and ALP lifetime.
Depending on the ALP lifetime, its decay may be effectively prompt or may take place visibly away from the interaction point. Correspondingly, we studied the sensitivity of both a prompt search and a displaced-vertex search.

In the case of the prompt search, we recast the Belle search for $\tau$ decays into three charged leptons~\cite{Hayasaka:2010np} to derive constraints on the couplings in the scenarios of interest.
The displaced search involves identifying the displaced vertex originating from the long-lived ALP decay.
In both cases we used simulated events to determine the efficiency, accounting for the size of the Belle~II tracker and for the dependence of the efficiency on the ALP decay position. 

We focused on two benchmark scenarios for detailed numerical results: Scenario 1 ($g_{\tau e}$ and $g_{ee}$ are non-zero) and Scenario 2 ($g_{\tau \mu}$ and $g_{\mu \mu}$ are non-zero), taking all other couplings to be vanishing.
The signatures for the two scenarios are, respectively,
$\tau \to X e, X\to e^-  e^+$ and $\tau\to X \mu, X\to  \mu^- \mu^+$.
Since the prompt Belle search was background-free~\cite{Hayasaka:2010np}, we expect this to also be the case at Belle~II, particularly with the displaced search. Therefore, we present 3-signal-event isocurves to indicate the Belle~II sensitivity at $95\%$ C.L. exclusion limits.
As a function of the ALP mass, we show both model-independent limits on the branching-fraction product BR$(\tau\to X e)\cdot$BR$(X\to e^- e^+)$ vs.~$c\tau_X$ and model-dependent limits on the ALP couplings.
Generally, the displaced-vertex (prompt) search shows better sensitivity for small (large)
values of $g_{\alpha \alpha}$, corresponding to long (short) decay lengths of the ALP.
For long decay lengths, the displaced-vertex search can extend the prompt search's sensitivity to the branching-fraction product by a factor of about 40, corresponding to about a factor of 6 in $g_{\tau l}$. 

The model-dependent results are also presented, for the case of equal production
and decay couplings, in the plane
$g=g_{\tau \alpha}=g_{\alpha \alpha}$ vs.~$m_X$.
In this case, in Scenario~1, the proposed displaced search is about as sensitive as constraints obtained from lepton-flavor universality tests in $\tau\to l\nu\bar\nu$.
However, in Scenario~2, the displaced search is sensitive to coupling values that are more than an order of magnitude smaller than those of the lepton-universality constraints.

Generally, we find that Belle II can probe
the LFV coupling $g_{\tau \alpha}$ down to below $10^{-10}$ GeV$^{-1}$, with the best sensitivity reached for a different value of $g_{\alpha \alpha}$ in each scenario.

%% file: alp_lfv_belleII.bbl
\providecommand{\href}[2]{#2}\begingroup\raggedright\begin{thebibliography}{100}

\bibitem{nEDM:2020crw}
{\bf nEDM} Collaboration, C.~Abel et~al., {\it {Measurement of the permanent
  electric dipole moment of the neutron}},  {\em Phys. Rev. Lett.} {\bf 124}
  (2020), no.~8 081803, [\href{http://arxiv.org/abs/2001.11966}{{\tt
  arXiv:2001.11966}}].

\bibitem{Dragos:2019oxn}
J.~Dragos, T.~Luu, A.~Shindler, J.~de~Vries, and A.~Yousif, {\it {Confirming
  the Existence of the strong CP Problem in Lattice QCD with the Gradient
  Flow}},  {\em Phys. Rev. C} {\bf 103} (2021), no.~1 015202,
  [\href{http://arxiv.org/abs/1902.03254}{{\tt arXiv:1902.03254}}].

\bibitem{Peccei:1977ur}
R.~D. Peccei and H.~R. Quinn, {\it {Constraints Imposed by CP Conservation in
  the Presence of Instantons}},  {\em Phys. Rev. D} {\bf 16} (1977) 1791--1797.

\bibitem{Peccei:2006as}
R.~D. Peccei, {\it {The Strong CP problem and axions}},  {\em Lect. Notes
  Phys.} {\bf 741} (2008) 3--17,
  [\href{http://arxiv.org/abs/hep-ph/0607268}{{\tt hep-ph/0607268}}].

\bibitem{ParticleDataGroup:2020ssz}
{\bf Particle Data Group} Collaboration, P.~A. Zyla et~al., {\it {Review of
  Particle Physics}},  {\em PTEP} {\bf 2020} (2020), no.~8 083C01.

\bibitem{Feng:1997tn}
J.~L. Feng, T.~Moroi, H.~Murayama, and E.~Schnapka, {\it {Third generation
  familons, b factories, and neutrino cosmology}},  {\em Phys. Rev. D} {\bf 57}
  (1998) 5875--5892, [\href{http://arxiv.org/abs/hep-ph/9709411}{{\tt
  hep-ph/9709411}}].

\bibitem{Kim:2015yna}
J.~E. Kim and D.~J.~E. Marsh, {\it {An ultralight pseudoscalar boson}},  {\em
  Phys. Rev. D} {\bf 93} (2016), no.~2 025027,
  [\href{http://arxiv.org/abs/1510.01701}{{\tt arXiv:1510.01701}}].

\bibitem{DeMartino:2017qsa}
I.~De~Martino, T.~Broadhurst, S.~H. Henry~Tye, T.~Chiueh, H.-Y. Schive, and
  R.~Lazkoz, {\it {Recognizing Axionic Dark Matter by Compton and de Broglie
  Scale Modulation of Pulsar Timing}},  {\em Phys. Rev. Lett.} {\bf 119}
  (2017), no.~22 221103, [\href{http://arxiv.org/abs/1705.04367}{{\tt
  arXiv:1705.04367}}].

\bibitem{Rubakov:1997vp}
V.~A. Rubakov, {\it {Grand unification and heavy axion}},  {\em JETP Lett.}
  {\bf 65} (1997) 621--624, [\href{http://arxiv.org/abs/hep-ph/9703409}{{\tt
  hep-ph/9703409}}].

\bibitem{Dine:1982ah}
M.~Dine and W.~Fischler, {\it {The Not So Harmless Axion}},  {\em Phys. Lett.
  B} {\bf 120} (1983) 137--141.

\bibitem{Abbott:1982af}
L.~F. Abbott and P.~Sikivie, {\it {A Cosmological Bound on the Invisible
  Axion}},  {\em Phys. Lett. B} {\bf 120} (1983) 133--136.

\bibitem{Preskill:1982cy}
J.~Preskill, M.~B. Wise, and F.~Wilczek, {\it {Cosmology of the Invisible
  Axion}},  {\em Phys. Lett. B} {\bf 120} (1983) 127--132.

\bibitem{Marsh:2015xka}
D.~J.~E. Marsh, {\it {Axion Cosmology}},  {\em Phys. Rept.} {\bf 643} (2016)
  1--79, [\href{http://arxiv.org/abs/1510.07633}{{\tt arXiv:1510.07633}}].

\bibitem{Lambiase:2018lhs}
G.~Lambiase and S.~Mohanty, {\it {Hydrogen spin oscillations in a background of
  axions and the 21-cm brightness temperature}},  {\em Mon. Not. Roy. Astron.
  Soc.} {\bf 494} (2020), no.~4 5961--5966,
  [\href{http://arxiv.org/abs/1804.05318}{{\tt arXiv:1804.05318}}].

\bibitem{Auriol:2018ovo}
A.~Auriol, S.~Davidson, and G.~Raffelt, {\it {Axion absorption and the spin
  temperature of primordial hydrogen}},  {\em Phys. Rev. D} {\bf 99} (2019),
  no.~2 023013, [\href{http://arxiv.org/abs/1808.09456}{{\tt
  arXiv:1808.09456}}].

\bibitem{Houston:2018vrf}
N.~Houston, C.~Li, T.~Li, Q.~Yang, and X.~Zhang, {\it {Natural Explanation for
  21 cm Absorption Signals via Axion-Induced Cooling}},  {\em Phys. Rev. Lett.}
  {\bf 121} (2018), no.~11 111301, [\href{http://arxiv.org/abs/1805.04426}{{\tt
  arXiv:1805.04426}}].

\bibitem{Jaeckel:2015jla}
J.~Jaeckel and M.~Spannowsky, {\it {Probing MeV to 90 GeV axion-like particles
  with LEP and LHC}},  {\em Phys. Lett. B} {\bf 753} (2016) 482--487,
  [\href{http://arxiv.org/abs/1509.00476}{{\tt arXiv:1509.00476}}].

\bibitem{Brivio:2017ije}
I.~Brivio, M.~B. Gavela, L.~Merlo, K.~Mimasu, J.~M. No, R.~del Rey, and
  V.~Sanz, {\it {ALPs Effective Field Theory and Collider Signatures}},  {\em
  Eur. Phys. J. C} {\bf 77} (2017), no.~8 572,
  [\href{http://arxiv.org/abs/1701.05379}{{\tt arXiv:1701.05379}}].

\bibitem{Dolan:2017osp}
M.~J. Dolan, T.~Ferber, C.~Hearty, F.~Kahlhoefer, and K.~Schmidt-Hoberg, {\it
  {Revised constraints and Belle II sensitivity for visible and invisible
  axion-like particles}},  {\em JHEP} {\bf 12} (2017) 094,
  [\href{http://arxiv.org/abs/1709.00009}{{\tt arXiv:1709.00009}}]. [Erratum:
  JHEP 03, 190 (2021)].

\bibitem{Bellazzini:2017neg}
B.~Bellazzini, A.~Mariotti, D.~Redigolo, F.~Sala, and J.~Serra, {\it {$R$-axion
  at colliders}},  {\em Phys. Rev. Lett.} {\bf 119} (2017), no.~14 141804,
  [\href{http://arxiv.org/abs/1702.02152}{{\tt arXiv:1702.02152}}].

\bibitem{Bauer:2017ris}
M.~Bauer, M.~Neubert, and A.~Thamm, {\it {Collider Probes of Axion-Like
  Particles}},  {\em JHEP} {\bf 12} (2017) 044,
  [\href{http://arxiv.org/abs/1708.00443}{{\tt arXiv:1708.00443}}].

\bibitem{Knapen:2017ebd}
S.~Knapen, T.~Lin, H.~K. Lou, and T.~Melia, {\it {LHC limits on axion-like
  particles from heavy-ion collisions}},  {\em CERN Proc.} {\bf 1} (2018) 65,
  [\href{http://arxiv.org/abs/1709.07110}{{\tt arXiv:1709.07110}}].

\bibitem{Bauer:2018uxu}
M.~Bauer, M.~Heiles, M.~Neubert, and A.~Thamm, {\it {Axion-Like Particles at
  Future Colliders}},  {\em Eur. Phys. J. C} {\bf 79} (2019), no.~1 74,
  [\href{http://arxiv.org/abs/1808.10323}{{\tt arXiv:1808.10323}}].

\bibitem{Aloni:2018vki}
D.~Aloni, Y.~Soreq, and M.~Williams, {\it {Coupling QCD-Scale Axionlike
  Particles to Gluons}},  {\em Phys. Rev. Lett.} {\bf 123} (2019), no.~3
  031803, [\href{http://arxiv.org/abs/1811.03474}{{\tt arXiv:1811.03474}}].

\bibitem{Carmona:2021seb}
A.~Carmona, C.~Scherb, and P.~Schwaller, {\it {Charming ALPs}},
  \href{http://arxiv.org/abs/2101.07803}{{\tt arXiv:2101.07803}}.

\bibitem{Belle-II:2020jti}
{\bf Belle-II} Collaboration, F.~Abudin\'en et~al., {\it {Search for Axion-Like
  Particles produced in $e^+e^-$ collisions at Belle II}},  {\em Phys. Rev.
  Lett.} {\bf 125} (2020), no.~16 161806,
  [\href{http://arxiv.org/abs/2007.13071}{{\tt arXiv:2007.13071}}].

\bibitem{Heeck:2016xwg}
J.~Heeck, {\it {Interpretation of Lepton Flavor Violation}},  {\em Phys. Rev.
  D} {\bf 95} (2017), no.~1 015022,
  [\href{http://arxiv.org/abs/1610.07623}{{\tt arXiv:1610.07623}}].

\bibitem{Petcov:1976ff}
S.~T. Petcov, {\it {The Processes $\mu \rightarrow e + \gamma, \mu \rightarrow
  e + \overline{e}, \nu' \rightarrow \nu + \gamma$ in the Weinberg-Salam Model
  with Neutrino Mixing}},  {\em Sov. J. Nucl. Phys.} {\bf 25} (1977) 340.
  [Erratum: Sov.J.Nucl.Phys. 25, 698 (1977), Erratum: Yad.Fiz. 25, 1336
  (1977)].

\bibitem{Hernandez-Tome:2018fbq}
G.~Hern\'andez-Tom\'e, G.~L\'opez~Castro, and P.~Roig, {\it {Flavor violating
  leptonic decays of $\tau$ and $\mu$ leptons in the Standard Model with
  massive neutrinos}},  {\em Eur. Phys. J. C} {\bf 79} (2019), no.~1 84,
  [\href{http://arxiv.org/abs/1807.06050}{{\tt arXiv:1807.06050}}]. [Erratum:
  Eur.Phys.J.C 80, 438 (2020)].

\bibitem{MEG:2016leq}
{\bf MEG} Collaboration, A.~M. Baldini et~al., {\it {Search for the lepton
  flavour violating decay $\mu ^+ \rightarrow \mathrm {e}^+ \gamma $ with the
  full dataset of the MEG experiment}},  {\em Eur. Phys. J. C} {\bf 76} (2016),
  no.~8 434, [\href{http://arxiv.org/abs/1605.05081}{{\tt arXiv:1605.05081}}].

\bibitem{SINDRUM:1987nra}
{\bf SINDRUM} Collaboration, U.~Bellgardt et~al., {\it {Search for the Decay
  mu+ ---\ensuremath{>} e+ e+ e-}},  {\em Nucl. Phys. B} {\bf 299} (1988) 1--6.

\bibitem{Blondel:2013ia}
A.~Blondel et~al., {\it {Research Proposal for an Experiment to Search for the
  Decay $\mu \to eee$}},  \href{http://arxiv.org/abs/1301.6113}{{\tt
  arXiv:1301.6113}}.

\bibitem{Baldini:2013ke}
A.~M. Baldini et~al., {\it {MEG Upgrade Proposal}},
  \href{http://arxiv.org/abs/1301.7225}{{\tt arXiv:1301.7225}}.

\bibitem{Cordero-Cid:2005vca}
A.~Cordero-Cid, G.~Tavares-Velasco, and J.~J. Toscano, {\it {Implications of a
  very light pseudoscalar boson on lepton flavor violation}},  {\em Phys. Rev.
  D} {\bf 72} (2005) 117701, [\href{http://arxiv.org/abs/hep-ph/0511331}{{\tt
  hep-ph/0511331}}].

\bibitem{Dev:2017ftk}
P.~S.~B. Dev, R.~N. Mohapatra, and Y.~Zhang, {\it {Lepton Flavor Violation
  Induced by a Neutral Scalar at Future Lepton Colliders}},  {\em Phys. Rev.
  Lett.} {\bf 120} (2018), no.~22 221804,
  [\href{http://arxiv.org/abs/1711.08430}{{\tt arXiv:1711.08430}}].

\bibitem{Bauer:2019gfk}
M.~Bauer, M.~Neubert, S.~Renner, M.~Schnubel, and A.~Thamm, {\it {Axionlike
  Particles, Lepton-Flavor Violation, and a New Explanation of $a_\mu$ and
  $a_e$}},  {\em Phys. Rev. Lett.} {\bf 124} (2020), no.~21 211803,
  [\href{http://arxiv.org/abs/1908.00008}{{\tt arXiv:1908.00008}}].

\bibitem{Cornella:2019uxs}
C.~Cornella, P.~Paradisi, and O.~Sumensari, {\it {Hunting for ALPs with Lepton
  Flavor Violation}},  {\em JHEP} {\bf 01} (2020) 158,
  [\href{http://arxiv.org/abs/1911.06279}{{\tt arXiv:1911.06279}}].

\bibitem{Calibbi:2020jvd}
L.~Calibbi, D.~Redigolo, R.~Ziegler, and J.~Zupan, {\it {Looking forward to
  Lepton-flavor-violating ALPs}},  \href{http://arxiv.org/abs/2006.04795}{{\tt
  arXiv:2006.04795}}.

\bibitem{Endo:2020mev}
M.~Endo, S.~Iguro, and T.~Kitahara, {\it {Probing $e\mu$ flavor-violating ALP
  at Belle II}},  {\em JHEP} {\bf 06} (2020) 040,
  [\href{http://arxiv.org/abs/2002.05948}{{\tt arXiv:2002.05948}}].

\bibitem{Batell:2009jf}
B.~Batell, M.~Pospelov, and A.~Ritz, {\it {Multi-lepton Signatures of a Hidden
  Sector in Rare B Decays}},  {\em Phys. Rev. D} {\bf 83} (2011) 054005,
  [\href{http://arxiv.org/abs/0911.4938}{{\tt arXiv:0911.4938}}].

\bibitem{Kamenik:2011vy}
J.~F. Kamenik and C.~Smith, {\it {FCNC portals to the dark sector}},  {\em
  JHEP} {\bf 03} (2012) 090, [\href{http://arxiv.org/abs/1111.6402}{{\tt
  arXiv:1111.6402}}].

\bibitem{Gavela:2019wzg}
M.~B. Gavela, R.~Houtz, P.~Quilez, R.~Del~Rey, and O.~Sumensari, {\it {Flavor
  constraints on electroweak ALP couplings}},  {\em Eur. Phys. J. C} {\bf 79}
  (2019), no.~5 369, [\href{http://arxiv.org/abs/1901.02031}{{\tt
  arXiv:1901.02031}}].

\bibitem{Belle-II:2010dht}
{\bf Belle-II} Collaboration, T.~Abe et~al., {\it {Belle II Technical Design
  Report}},  \href{http://arxiv.org/abs/1011.0352}{{\tt arXiv:1011.0352}}.

\bibitem{Belle-II:2018jsg}
{\bf Belle-II} Collaboration, W.~Altmannshofer et~al., {\it {The Belle II
  Physics Book}},  {\em PTEP} {\bf 2019} (2019), no.~12 123C01,
  [\href{http://arxiv.org/abs/1808.10567}{{\tt arXiv:1808.10567}}]. [Erratum:
  PTEP 2020, 029201 (2020)].

\bibitem{HernandezVillanueva:2018dqu}
{\bf Belle-II} Collaboration, M.~Hern\'andez~Villanueva, {\it {Prospects for
  $\tau$ lepton physics at Belle II}},  {\em SciPost Phys. Proc.} {\bf 1}
  (2019) 003, [\href{http://arxiv.org/abs/1812.04225}{{\tt arXiv:1812.04225}}].

\bibitem{Essig:2010gu}
R.~Essig, R.~Harnik, J.~Kaplan, and N.~Toro, {\it {Discovering New Light States
  at Neutrino Experiments}},  {\em Phys. Rev. D} {\bf 82} (2010) 113008,
  [\href{http://arxiv.org/abs/1008.0636}{{\tt arXiv:1008.0636}}].

\bibitem{Andreas:2010ms}
S.~Andreas, O.~Lebedev, S.~Ramos-Sanchez, and A.~Ringwald, {\it {Constraints on
  a very light CP-odd Higgs of the NMSSM and other axion-like particles}},
  {\em JHEP} {\bf 08} (2010) 003, [\href{http://arxiv.org/abs/1005.3978}{{\tt
  arXiv:1005.3978}}].

\bibitem{BaBar:2016sci}
{\bf BaBar} Collaboration, J.~P. Lees et~al., {\it {Search for a muonic dark
  force at BABAR}},  {\em Phys. Rev. D} {\bf 94} (2016), no.~1 011102,
  [\href{http://arxiv.org/abs/1606.03501}{{\tt arXiv:1606.03501}}].

\bibitem{ARGUS:1995bjh}
{\bf ARGUS} Collaboration, H.~Albrecht et~al., {\it {A Search for lepton flavor
  violating decays tau ----\ensuremath{>} e alpha, tau ---\ensuremath{>} mu
  alpha}},  {\em Z. Phys. C} {\bf 68} (1995) 25--28.

\bibitem{Yoshinobu:2017jti}
{\bf Belle} Collaboration, T.~Yoshinobu and K.~Hayasaka, {\it {MC study for the
  lepton flavor violating tau decay into a lepton and an undetectable
  particle}},  {\em Nucl. Part. Phys. Proc.} {\bf 287-288} (2017) 218--220.

\bibitem{Guadagnoli:2021fcj}
D.~Guadagnoli, C.~B. Park, and F.~Tenchini, {\it {$\tau \to \ell +$ invisible
  through invisible-savvy collider variables}},
  \href{http://arxiv.org/abs/2106.16236}{{\tt arXiv:2106.16236}}.

\bibitem{DeLaCruz-Burelo:2020ozf}
E.~De~La Cruz-Burelo, M.~Hernandez-Villanueva, and A.~De~Yta-Hernandez, {\it
  {New method for beyond the Standard Model invisible particle searches in tau
  lepton decays}},  {\em Phys. Rev. D} {\bf 102} (2020), no.~11 115001,
  [\href{http://arxiv.org/abs/2007.08239}{{\tt arXiv:2007.08239}}].

\bibitem{Ma:2021jkp}
K.~Ma, {\it {Polarization Effects in Lepton Flavor Violated Decays Induced by
  Axion-Like Particles}},  \href{http://arxiv.org/abs/2104.11162}{{\tt
  arXiv:2104.11162}}.

\bibitem{Tenchini:2020njf}
F.~Tenchini, M.~Garcia-Hernandez, T.~Kraetzschmar, P.~K. Rados, E.~De~La
  Cruz-Burelo, A.~De~Yta-Hernandez, I.~Heredia de~la Cruz, and A.~Rostomyan,
  {\it {First results and prospects for tau LFV decay $\tau \rightarrow e +
  \alpha$(invisible) at Belle II}},  {\em PoS} {\bf ICHEP2020} (2021) 288.

\bibitem{BaBar:2009lyd}
{\bf BaBar} Collaboration, B.~Aubert et~al., {\it {Measurements of Charged
  Current Lepton Universality and $|V_{us}|$ using Tau Lepton Decays to $e^-
  \bar{\nu}_e \nu_\tau$, $\mu^- \bar{\nu}_\mu \nu_\tau$, $\pi^- \nu_\tau$, and
  $K^- \nu_\tau$}},  {\em Phys. Rev. Lett.} {\bf 105} (2010) 051602,
  [\href{http://arxiv.org/abs/0912.0242}{{\tt arXiv:0912.0242}}].

\bibitem{Iguro:2020rby}
S.~Iguro, Y.~Omura, and M.~Takeuchi, {\it {Probing $\mu\tau$ flavor-violating
  solutions for the muon $g-2$ anomaly at Belle II}},  {\em JHEP} {\bf 09}
  (2020) 144, [\href{http://arxiv.org/abs/2002.12728}{{\tt arXiv:2002.12728}}].

\bibitem{Belle:2000cnh}
{\bf Belle} Collaboration, A.~Abashian et~al., {\it {The Belle Detector}},
  {\em Nucl. Instrum. Meth. A} {\bf 479} (2002) 117--232.

\bibitem{Hayasaka:2010np}
K.~Hayasaka et~al., {\it {Search for Lepton Flavor Violating Tau Decays into
  Three Leptons with 719 Million Produced Tau+Tau- Pairs}},  {\em Phys. Lett.
  B} {\bf 687} (2010) 139--143, [\href{http://arxiv.org/abs/1001.3221}{{\tt
  arXiv:1001.3221}}].

\bibitem{BaBar:2010axs}
{\bf BaBar} Collaboration, J.~P. Lees et~al., {\it {Limits on tau Lepton-Flavor
  Violating Decays in three charged leptons}},  {\em Phys. Rev. D} {\bf 81}
  (2010) 111101, [\href{http://arxiv.org/abs/1002.4550}{{\tt
  arXiv:1002.4550}}].

\bibitem{Curtin:2018mvb}
D.~Curtin et~al., {\it {Long-Lived Particles at the Energy Frontier: The
  MATHUSLA Physics Case}},  {\em Rept. Prog. Phys.} {\bf 82} (2019), no.~11
  116201, [\href{http://arxiv.org/abs/1806.07396}{{\tt arXiv:1806.07396}}].

\bibitem{Lee:2018pag}
L.~Lee, C.~Ohm, A.~Soffer, and T.-T. Yu, {\it {Collider Searches for Long-Lived
  Particles Beyond the Standard Model}},  {\em Prog. Part. Nucl. Phys.} {\bf
  106} (2019) 210--255, [\href{http://arxiv.org/abs/1810.12602}{{\tt
  arXiv:1810.12602}}].

\bibitem{Alimena:2019zri}
J.~Alimena et~al., {\it {Searching for long-lived particles beyond the Standard
  Model at the Large Hadron Collider}},  {\em J. Phys. G} {\bf 47} (2020),
  no.~9 090501, [\href{http://arxiv.org/abs/1903.04497}{{\tt
  arXiv:1903.04497}}].

\bibitem{Dib:2019tuj}
C.~O. Dib, J.~C. Helo, M.~Nayak, N.~A. Neill, A.~Soffer, and J.~Zamora-Saa,
  {\it {Searching for a sterile neutrino that mixes predominantly with
  $\nu_\tau$ at $B$ factories}},  {\em Phys. Rev. D} {\bf 101} (2020), no.~9
  093003, [\href{http://arxiv.org/abs/1908.09719}{{\tt arXiv:1908.09719}}].

\bibitem{Kim:2019xqj}
C.~S. Kim, Y.~Kwon, D.~Lee, S.~Oh, and D.~Sahoo, {\it {Probing sterile
  neutrinos in $B (D)$ meson decays at Belle II (BESIII)}},  {\em Eur. Phys. J.
  C} {\bf 80} (2020), no.~8 730, [\href{http://arxiv.org/abs/1908.00376}{{\tt
  arXiv:1908.00376}}].

\bibitem{Kang:2021oes}
D.~W. Kang, P.~Ko, and C.-T. Lu, {\it {Exploring properties of long-lived
  particles in inelastic dark matter models at Belle II}},  {\em JHEP} {\bf 04}
  (2021) 269, [\href{http://arxiv.org/abs/2101.02503}{{\tt arXiv:2101.02503}}].

\bibitem{Acevedo:2021wiq}
M.~Acevedo, A.~Blackburn, N.~Blinov, B.~Shuve, and M.~Stone, {\it {Multi-track
  Displaced Vertices at B-Factories}},
  \href{http://arxiv.org/abs/2105.12744}{{\tt arXiv:2105.12744}}.

\bibitem{Dreyer:2021aqd}
S.~Dreyer et~al., {\it {Physics reach of a long-lived particle detector at
  Belle II}},  \href{http://arxiv.org/abs/2105.12962}{{\tt arXiv:2105.12962}}.

\bibitem{Duerr:2019dmv}
M.~Duerr, T.~Ferber, C.~Hearty, F.~Kahlhoefer, K.~Schmidt-Hoberg, and
  P.~Tunney, {\it {Invisible and displaced dark matter signatures at Belle
  II}},  {\em JHEP} {\bf 02} (2020) 039,
  [\href{http://arxiv.org/abs/1911.03176}{{\tt arXiv:1911.03176}}].

\bibitem{Duerr:2020muu}
M.~Duerr, T.~Ferber, C.~Garcia-Cely, C.~Hearty, and K.~Schmidt-Hoberg, {\it
  {Long-lived Dark Higgs and Inelastic Dark Matter at Belle II}},  {\em JHEP}
  {\bf 04} (2021) 146, [\href{http://arxiv.org/abs/2012.08595}{{\tt
  arXiv:2012.08595}}].

\bibitem{Filimonova:2019tuy}
A.~Filimonova, R.~Sch\"afer, and S.~Westhoff, {\it {Probing dark sectors with
  long-lived particles at BELLE II}},  {\em Phys. Rev. D} {\bf 101} (2020),
  no.~9 095006, [\href{http://arxiv.org/abs/1911.03490}{{\tt
  arXiv:1911.03490}}].

\bibitem{Chen:2020bok}
X.~Chen, Z.~Hu, Y.~Wu, and K.~Yi, {\it {Search for dark photon and dark matter
  signatures around electron-positron colliders}},  {\em Phys. Lett. B} {\bf
  814} (2021) 136076, [\href{http://arxiv.org/abs/2001.04382}{{\tt
  arXiv:2001.04382}}].

\bibitem{Dey:2020juy}
S.~Dey, C.~O. Dib, J.~Carlos~Helo, M.~Nayak, N.~A. Neill, A.~Soffer, and Z.~S.
  Wang, {\it {Long-lived light neutralinos at Belle II}},  {\em JHEP} {\bf 02}
  (2021) 211, [\href{http://arxiv.org/abs/2012.00438}{{\tt arXiv:2012.00438}}].

\bibitem{Bertholet:2021hjl}
E.~Bertholet, S.~Chakraborty, V.~Loladze, T.~Okui, A.~Soffer, and K.~Tobioka,
  {\it {Heavy QCD Axion at Belle II: Displaced and Prompt Signals}},
  \href{http://arxiv.org/abs/2108.10331}{{\tt arXiv:2108.10331}}.

\bibitem{Belle:2013ytx}
{\bf Belle} Collaboration, D.~Liventsev et~al., {\it {Search for heavy
  neutrinos at Belle}},  {\em Phys. Rev. D} {\bf 87} (2013), no.~7 071102,
  [\href{http://arxiv.org/abs/1301.1105}{{\tt arXiv:1301.1105}}]. [Erratum:
  Phys.Rev.D 95, 099903 (2017)].

\bibitem{BaBar:2015jvu}
{\bf BaBar} Collaboration, J.~P. Lees et~al., {\it {Search for Long-Lived
  Particles in $e^+e^-$ Collisions}},  {\em Phys. Rev. Lett.} {\bf 114} (2015),
  no.~17 171801, [\href{http://arxiv.org/abs/1502.02580}{{\tt
  arXiv:1502.02580}}].

\bibitem{Heeck:2017xmg}
J.~Heeck and W.~Rodejohann, {\it {Lepton flavor violation with displaced
  vertices}},  {\em Phys. Lett. B} {\bf 776} (2018) 385--390,
  [\href{http://arxiv.org/abs/1710.02062}{{\tt arXiv:1710.02062}}].

\bibitem{ALEPH:2006jhv}
{\bf ALEPH} Collaboration, S.~Schael et~al., {\it {Fermion pair production in
  $e^{+} e^{-}$ collisions at 189-209-GeV and constraints on physics beyond the
  standard model}},  {\em Eur. Phys. J. C} {\bf 49} (2007) 411--437,
  [\href{http://arxiv.org/abs/hep-ex/0609051}{{\tt hep-ex/0609051}}].

\bibitem{Christensen:2008py}
N.~D. Christensen and C.~Duhr, {\it {FeynRules - Feynman rules made easy}},
  {\em Comput. Phys. Commun.} {\bf 180} (2009) 1614--1641,
  [\href{http://arxiv.org/abs/0806.4194}{{\tt arXiv:0806.4194}}].

\bibitem{Alloul:2013bka}
A.~Alloul, N.~D. Christensen, C.~Degrande, C.~Duhr, and B.~Fuks, {\it
  {FeynRules 2.0 - A complete toolbox for tree-level phenomenology}},  {\em
  Comput. Phys. Commun.} {\bf 185} (2014) 2250--2300,
  [\href{http://arxiv.org/abs/1310.1921}{{\tt arXiv:1310.1921}}].

\bibitem{Degrande:2011ua}
C.~Degrande, C.~Duhr, B.~Fuks, D.~Grellscheid, O.~Mattelaer, and T.~Reiter,
  {\it {UFO - The Universal FeynRules Output}},  {\em Comput. Phys. Commun.}
  {\bf 183} (2012) 1201--1214, [\href{http://arxiv.org/abs/1108.2040}{{\tt
  arXiv:1108.2040}}].

\bibitem{Alwall:2014hca}
J.~Alwall, R.~Frederix, S.~Frixione, V.~Hirschi, F.~Maltoni, O.~Mattelaer,
  H.~S. Shao, T.~Stelzer, P.~Torrielli, and M.~Zaro, {\it {The automated
  computation of tree-level and next-to-leading order differential cross
  sections, and their matching to parton shower simulations}},  {\em JHEP} {\bf
  07} (2014) 079, [\href{http://arxiv.org/abs/1405.0301}{{\tt
  arXiv:1405.0301}}].

\bibitem{SLACE158:2005uay}
{\bf SLAC E158} Collaboration, P.~L. Anthony et~al., {\it {Precision
  measurement of the weak mixing angle in Moller scattering}},  {\em Phys. Rev.
  Lett.} {\bf 95} (2005) 081601,
  [\href{http://arxiv.org/abs/hep-ex/0504049}{{\tt hep-ex/0504049}}].

\bibitem{Hallin:1992sj}
A.~L. Hallin, F.~P. Calaprice, R.~A. McPherson, and E.~R.~J. Saettler, {\it
  {Sensitive search for resonances in low-energy e+ e- scattering}},  {\em
  Phys. Rev. D} {\bf 45} (1992) 3955--3960.

\bibitem{Haghighat:2021djz}
G.~Haghighat and M.~Mohammadi~Najafabadi, {\it {Search for
  lepton-flavor-violating ALPs at a future muon collider and utilization of
  polarization-induced effects}},  \href{http://arxiv.org/abs/2106.00505}{{\tt
  arXiv:2106.00505}}.

\bibitem{Delahaye:2019omf}
J.~P. Delahaye, M.~Diemoz, K.~Long, B.~Mansouli\'e, N.~Pastrone, L.~Rivkin,
  D.~Schulte, A.~Skrinsky, and A.~Wulzer, {\it {Muon Colliders}},
  \href{http://arxiv.org/abs/1901.06150}{{\tt arXiv:1901.06150}}.

\bibitem{Boscolo:2018ytm}
M.~Boscolo, J.-P. Delahaye, and M.~Palmer, {\it {The future prospects of muon
  colliders and neutrino factories}},  {\em Rev. Accel. Sci. Tech.} {\bf 10}
  (2019), no.~01 189--214, [\href{http://arxiv.org/abs/1808.01858}{{\tt
  arXiv:1808.01858}}].

\bibitem{AlAli:2021let}
H.~Al~Ali et~al., {\it {The Muon Smasher's Guide}},
  \href{http://arxiv.org/abs/2103.14043}{{\tt arXiv:2103.14043}}.

\bibitem{Bollig:2020xdr}
R.~Bollig, W.~DeRocco, P.~W. Graham, and H.-T. Janka, {\it {Muons in
  Supernovae: Implications for the Axion-Muon Coupling}},  {\em Phys. Rev.
  Lett.} {\bf 125} (2020), no.~5 051104,
  [\href{http://arxiv.org/abs/2005.07141}{{\tt arXiv:2005.07141}}]. [Erratum:
  Phys.Rev.Lett. 126, 189901 (2021)].

\bibitem{Lucente:2021hbp}
G.~Lucente and P.~Carenza, {\it {Supernova bound on Axion-Like Particles
  coupled with electrons}},  \href{http://arxiv.org/abs/2107.12393}{{\tt
  arXiv:2107.12393}}.

\bibitem{Foldenauer:2016rpi}
P.~Foldenauer and J.~Jaeckel, {\it {Purely flavor-changing Z' bosons and where
  they might hide}},  {\em JHEP} {\bf 05} (2017) 010,
  [\href{http://arxiv.org/abs/1612.07789}{{\tt arXiv:1612.07789}}].

\bibitem{Altmannshofer:2016brv}
W.~Altmannshofer, C.-Y. Chen, P.~S. Bhupal~Dev, and A.~Soni, {\it {Lepton
  flavor violating Z' explanation of the muon anomalous magnetic moment}},
  {\em Phys. Lett. B} {\bf 762} (2016) 389--398,
  [\href{http://arxiv.org/abs/1607.06832}{{\tt arXiv:1607.06832}}].

\bibitem{Pich:2013lsa}
A.~Pich, {\it {Precision Tau Physics}},  {\em Prog. Part. Nucl. Phys.} {\bf 75}
  (2014) 41--85, [\href{http://arxiv.org/abs/1310.7922}{{\tt
  arXiv:1310.7922}}].

\bibitem{Bryman:2021rtr}
D.~A. Bryman, S.~Ito, and R.~Shrock, {\it {Upper Limits on Branching Ratios of
  Decays $\tau \to \ell \gamma\gamma$ and $\tau \to \ell X$}},
  \href{http://arxiv.org/abs/2106.02451}{{\tt arXiv:2106.02451}}.

\bibitem{BaBar:2009hkt}
{\bf BaBar} Collaboration, B.~Aubert et~al., {\it {Searches for Lepton Flavor
  Violation in the Decays tau+- ---\ensuremath{>} e+- gamma and tau+-
  ---\ensuremath{>} mu+- gamma}},  {\em Phys. Rev. Lett.} {\bf 104} (2010)
  021802, [\href{http://arxiv.org/abs/0908.2381}{{\tt arXiv:0908.2381}}].

\bibitem{Morel:2020dww}
L.~Morel, Z.~Yao, P.~Clad\'e, and S.~Guellati-Kh\'elifa, {\it {Determination of
  the fine-structure constant with an accuracy of 81 parts per trillion}},
  {\em Nature} {\bf 588} (2020), no.~7836 61--65.

\bibitem{Muong-2:2021ojo}
{\bf Muon g-2} Collaboration, B.~Abi et~al., {\it {Measurement of the Positive
  Muon Anomalous Magnetic Moment to 0.46 ppm}},  {\em Phys. Rev. Lett.} {\bf
  126} (2021), no.~14 141801, [\href{http://arxiv.org/abs/2104.03281}{{\tt
  arXiv:2104.03281}}].

\bibitem{Muong-2:2006rrc}
{\bf Muon g-2} Collaboration, G.~W. Bennett et~al., {\it {Final Report of the
  Muon E821 Anomalous Magnetic Moment Measurement at BNL}},  {\em Phys. Rev. D}
  {\bf 73} (2006) 072003, [\href{http://arxiv.org/abs/hep-ex/0602035}{{\tt
  hep-ex/0602035}}].

\bibitem{Aoyama:2020ynm}
T.~Aoyama et~al., {\it {The anomalous magnetic moment of the muon in the
  Standard Model}},  {\em Phys. Rept.} {\bf 887} (2020) 1--166,
  [\href{http://arxiv.org/abs/2006.04822}{{\tt arXiv:2006.04822}}].

\bibitem{Buen-Abad:2021fwq}
M.~A. Buen-Abad, J.~Fan, M.~Reece, and C.~Sun, {\it {Challenges for an axion
  explanation of the muon $g-2$ measurement}},
  \href{http://arxiv.org/abs/2104.03267}{{\tt arXiv:2104.03267}}.

\bibitem{Keung:2021rps}
W.-Y. Keung, D.~Marfatia, and P.-Y. Tseng, {\it {Axion-like particles,
  two-Higgs-doublet models, leptoquarks, and the electron and muon g-2}},  {\em
  LHEP} {\bf 2021} (2021) 209, [\href{http://arxiv.org/abs/2104.03341}{{\tt
  arXiv:2104.03341}}].

\bibitem{Borsanyi:2020mff}
S.~Borsanyi et~al., {\it {Leading hadronic contribution to the muon magnetic
  moment from lattice QCD}},  {\em Nature} {\bf 593} (2021), no.~7857 51--55,
  [\href{http://arxiv.org/abs/2002.12347}{{\tt arXiv:2002.12347}}].

\bibitem{Riordan:1987aw}
E.~M. Riordan et~al., {\it {A Search for Short Lived Axions in an Electron Beam
  Dump Experiment}},  {\em Phys. Rev. Lett.} {\bf 59} (1987) 755.

\bibitem{CHARM:1985anb}
{\bf CHARM} Collaboration, F.~Bergsma et~al., {\it {Search for Axion Like
  Particle Production in 400-{GeV} Proton - Copper Interactions}},  {\em Phys.
  Lett. B} {\bf 157} (1985) 458--462.

\bibitem{Davier:1989wz}
M.~Davier and H.~Nguyen~Ngoc, {\it {An Unambiguous Search for a Light Higgs
  Boson}},  {\em Phys. Lett. B} {\bf 229} (1989) 150--155.

\bibitem{Bross:1989mp}
A.~Bross, M.~Crisler, S.~H. Pordes, J.~Volk, S.~Errede, and J.~Wrbanek, {\it {A
  Search for Shortlived Particles Produced in an Electron Beam Dump}},  {\em
  Phys. Rev. Lett.} {\bf 67} (1991) 2942--2945.

\bibitem{Willmann:1998gd}
L.~Willmann et~al., {\it {New bounds from searching for muonium to anti-muonium
  conversion}},  {\em Phys. Rev. Lett.} {\bf 82} (1999) 49--52,
  [\href{http://arxiv.org/abs/hep-ex/9807011}{{\tt hep-ex/9807011}}].

\bibitem{Calibbi:2017uvl}
L.~Calibbi and G.~Signorelli, {\it {Charged Lepton Flavour Violation: An
  Experimental and Theoretical Introduction}},  {\em Riv. Nuovo Cim.} {\bf 41}
  (2018), no.~2 71--174, [\href{http://arxiv.org/abs/1709.00294}{{\tt
  arXiv:1709.00294}}].

\bibitem{SINDRUMII:2006dvw}
{\bf SINDRUM II} Collaboration, W.~H. Bertl et~al., {\it {A Search for muon to
  electron conversion in muonic gold}},  {\em Eur. Phys. J. C} {\bf 47} (2006)
  337--346.

\bibitem{Mu2e:2014fns}
{\bf Mu2e} Collaboration, L.~Bartoszek et~al., {\it {Mu2e Technical Design
  Report}},  \href{http://arxiv.org/abs/1501.05241}{{\tt arXiv:1501.05241}}.

\bibitem{COMET:2018auw}
{\bf COMET} Collaboration, R.~Abramishvili et~al., {\it {COMET Phase-I
  Technical Design Report}},  {\em PTEP} {\bf 2020} (2020), no.~3 033C01,
  [\href{http://arxiv.org/abs/1812.09018}{{\tt arXiv:1812.09018}}].

\bibitem{Sjostrand:2014zea}
T.~Sj\"ostrand, S.~Ask, J.~R. Christiansen, R.~Corke, N.~Desai, P.~Ilten,
  S.~Mrenna, S.~Prestel, C.~O. Rasmussen, and P.~Z. Skands, {\it {An
  introduction to PYTHIA 8.2}},  {\em Comput. Phys. Commun.} {\bf 191} (2015)
  159--177, [\href{http://arxiv.org/abs/1410.3012}{{\tt arXiv:1410.3012}}].

\bibitem{Heeck:2016xkh}
J.~Heeck, {\it {Lepton flavor violation with light vector bosons}},  {\em Phys.
  Lett. B} {\bf 758} (2016) 101--105,
  [\href{http://arxiv.org/abs/1602.03810}{{\tt arXiv:1602.03810}}].

\bibitem{Chen:2017cic}
C.-H. Chen and T.~Nomura, {\it {$L_\mu -L_\tau$ gauge-boson production from
  lepton flavor violating $\tau$ decays at Belle II}},  {\em Phys. Rev. D} {\bf
  96} (2017), no.~9 095023, [\href{http://arxiv.org/abs/1704.04407}{{\tt
  arXiv:1704.04407}}].

\bibitem{Flacke:2016szy}
T.~Flacke, C.~Frugiuele, E.~Fuchs, R.~S. Gupta, and G.~Perez, {\it
  {Phenomenology of relaxion-Higgs mixing}},  {\em JHEP} {\bf 06} (2017) 050,
  [\href{http://arxiv.org/abs/1610.02025}{{\tt arXiv:1610.02025}}].

\bibitem{BhupalDev:2016nfr}
P.~S. Bhupal~Dev, R.~N. Mohapatra, and Y.~Zhang, {\it {Displaced photon signal
  from a possible light scalar in minimal left-right seesaw model}},  {\em
  Phys. Rev. D} {\bf 95} (2017), no.~11 115001,
  [\href{http://arxiv.org/abs/1612.09587}{{\tt arXiv:1612.09587}}].

\bibitem{Dev:2017dui}
P.~S.~B. Dev, R.~N. Mohapatra, and Y.~Zhang, {\it {Long Lived Light Scalars as
  Probe of Low Scale Seesaw Models}},  {\em Nucl. Phys. B} {\bf 923} (2017)
  179--221, [\href{http://arxiv.org/abs/1703.02471}{{\tt arXiv:1703.02471}}].

\bibitem{BaBar:2020jma}
{\bf BaBar} Collaboration, J.~P. Lees et~al., {\it {Search for a Dark
  Leptophilic Scalar in $e^+e^-$ Collisions}},  {\em Phys. Rev. Lett.} {\bf
  125} (2020), no.~18 181801, [\href{http://arxiv.org/abs/2005.01885}{{\tt
  arXiv:2005.01885}}].

\end{thebibliography}\endgroup
